\documentclass[a4paper,preprint,superscriptaddress,preprintnumbers,nofootinbib]{revtex4}
\usepackage[dvipdfmx,hiresbb]{graphicx}
\usepackage{color}
\usepackage{amsmath}
\usepackage{amssymb}
\usepackage{wrapfig}
\usepackage{bm}

\allowdisplaybreaks

\catcode`\@=11
\def\lsim{\mathrel{\mathpalette\@versim<}}
\def\gsim{\mathrel{\mathpalette\@versim>}}
\def\@versim#1#2{\vcenter{\offinterlineskip
\ialign{$\m@th#1\hfil##\hfil$\crcr#2\crcr\sim\crcr } }}
\catcode`\@=12

\newcommand{\Slash}[1]{{\ooalign{\hfil/\hfil\crcr$#1$}}}

\newcommand{\SU}{\text{SU}}
\newcommand{\U}{\text{U}}

\newcommand{\al}[1]{\begin{align}#1\end{align}}
\newcommand{\bp}{\begin{pmatrix}}
\newcommand{\ep}{\end{pmatrix}}
\newcommand{\nn}{\nonumber\\}

\newcommand{\p}{\partial}

\newcommand{\df}{\text{d}}

\newcommand{\bs}[1]{\boldsymbol}

\newcommand{\pmat}[1]{\begin{pmatrix}#1\end{pmatrix}}

\newcommand{\fn}[1]{\!\left(#1\right)}
\newcommand{\bra}{\langle}
\newcommand{\ket}{\rangle}
\newcommand{\Lag}{\mathcal L}
\newcommand{\tr}{\text{tr}}

\begin{document}
\title{Gravitational wave from dark sector with dark pion}

\author{Koji \surname{Tsumura}}
\email{ko2@gauge.scphys.kyoto-u.ac.jp }
\affiliation{Department of Physics, Kyoto University, Kyoto 606-8502, Japan}

\author{Masatoshi \surname{Yamada}}
\email{m.yamada@thphys.uni-heidelberg.de}
\affiliation{Institut f\"ur Theoretische Physik, Universit\"at Heidelberg, Philosophenweg 16, 69120 Heidelberg, Germany
}

\author{Yuya \surname{Yamaguchi}}
\email{yy@particle.sci.hokudai.ac.jp}
\affiliation{Department of Physics, Faculty of Science, Hokkaido University, Sapporo 060-0810, Japan}

\preprint{
KUNS-2669
}
\preprint{
EPHOU-17-004
}
\begin{abstract}
In this work, we investigate the spectra of gravitational waves produced by chiral symmetry breaking in dark quantum chromodynamics (dQCD) sector.
The dark pion ($\pi$) can be a dark matter candidate as weakly interacting massive particle (WIMP) or strongly interacting massive particle (SIMP).
For a WIMP scenario, we introduce the dQCD sector coupled to the standard model (SM) sector with classical scale invariance and investigate the annihilation process of the dark pion via the $2\pi \to 2\,\text{SM}$ process.
For a SIMP scenario, we investigate the $3\pi \to 2\pi$ annihilation process of the dark pion as a SIMP using chiral perturbation theory.
We find that in the WIMP scenario the gravitational wave background spectra can be observed by future space gravitational wave antennas.
On the other hand, when the dark pion is the SIMP dark matter with the constraints for the chiral perturbative limit and pion-pion scattering cross section, the chiral phase transition becomes crossover and then the gravitational waves are not produced.
\end{abstract}
\maketitle

\newpage
\section{Introduction}
One of greatest unsolved mysteries in elementary particle physics is an evidence of  dark matter (DM).
The Planck observation tells us that its relic abundance today is $\Omega_\text{DM}h^2\simeq 0.1186\pm 0.02$~\cite{Ade:2015xua}.
There is no candidate of DM in the standard model (SM), and therefore it has to be extended.

One of attractive candidates of dark matter is the weakly interacting massive particle (WIMP),
whose thermal abundance is determined by 2\,DM\,$\to$\, 2\,SM annihilation process. 
Such a particle naturally emerges from an extension of the SM, e.g., supersymmetry.
Recently, a new scenario for DM, namely strongly interacting massive particle (SIMP), was suggested~\cite{Hochberg:2014dra}.
In this case, the annihilation process proceeds the 3\,DM\, $\to$\, 2\,DM process.
The SIMP model could naturally address the core-cusp problem~\cite{Hochberg:2014dra}.

As one of extensions of the SM, dark quantum chromodynamics (dQCD) has been often introduced~\cite{Hur:2007uz,Hur:2011sv,Heikinheimo:2013fta,Farzinnia:2013pga,Foot:2013hna}.
Since the dark pion is generally stable thanks to the flavor symmetry, it can be a dark matter candidate.
Within extensions of the Higgs sector with dQCD, the annihilation process 2$\pi$\,$\to$\, 2\,SM is naturally given via the Higgs portal coupling; see e.g.~\cite{Heikinheimo:2013fta,Farzinnia:2013pga,Holthausen:2013ota}.
In such a case, the dark pion has a TeV scale mass and then behaves as a WIMP dark matter.
It is also known that the dark pion can be a SIMP in dQCD~\cite{Hochberg:2014kqa,Hochberg:2015vrg}.
In the SIMP scenario, the Wess-Zumino-Witten term~\cite{Wess:1971yu,Witten:1983tw,Witten:1983tx} with the five point interaction plays a crucial role for the $3\pi\to2\pi$ process.

It is important to detect the signals from the dark pion by experiments such as collider experiments and the DM detections.
We need the complementary probes towards observing the dark pion in order to clarify the nature of DM.
As a new way of probing new physics, the observation of the spectra of gravitational waves (GWs) may be significant.
After the discovery of the GW~\cite{Abbott:2016blz}, it has been accelerated to study towards observing stochastic GWs produced by the inflation and the cosmic strings.
Several experiments have been designed, e.g. eLISA~\cite{Seoane:2013qna,Audley:2017drz} and DECIGO~\cite{Seto:2001qf,Kawamura:2006up,Kawamura:2011zz} towards observing the wide-band of frequency of the GWs.
The early studies~\cite{Witten:1984rs,Hogan:1986qda,Turner:1990rc} have pointed out that the cosmological phase transitions can produce the GWs.
Refs.~\cite{Maggiore:1999vm,Kahniashvili:2008pf,Caprini:2010xv,Binetruy:2012ze,Schwaller:2015tja,Caprini:2015zlo,Huber:2015znp,Leitao:2015fmj,Dev:2016feu,Addazi:2016fbj,Huang:2016cjm,Huang:2017laj,Cai:2017cbj,Addazi:2017gpt} have discussed the possibility of their observations in future experiments.

Let us consider a strongly interacting dark sector whose Lagrangian is described by
\al{
\Lag_\text{dQCD}=-\frac{1}{4}F_{\mu\nu}^aF^a{}^{\mu\nu} + \bar \psi_i (\delta _{ij}i\Slash D - (m_q)_{ij}) \psi_j,
\label{original Lag}
}
where $F_{\mu\nu}^a$ is the field strength for $\SU\fn{N_c}$ gauge fields $A_\mu^a$; $D_\mu=\p_\mu-ig_\text{H}A_\mu^aT^a$ is the covariant derivative; $N_c$ is the dark color number; and $\psi_i$ is the dark quark with a finite mass $(m_q)_{ij}$ and the flavor number $i,j=1,...,N_f$.
We assume that the dark quarks do not have the SM charge.
Note that if the dark quark masses are degenerate, i.e., $(m_q)_{ij}=m_q\delta_{ij}$, the Lagrangian Eq.~\eqref{original Lag} has global symmetries: $\text{SU}\fn{N_f}_{V}\times \text{U}\fn{1}_V\times \text{U}\fn{1}_A$.

As explained above, this dark pion can be a WIMP or a SIMP dark matter.
Besides it, one can easily expect that the phase transitions at finite temperature in the dark sector have taken place in the early universe due to the strong dynamics if the dark pion actually exists.
It is known that for the small quark masses for $N_f=3$, the chiral phase transition becomes strongly first-order, which can be seen in the Columbia plot, see e.g.~\cite{Brown:1990ev}.
In this case, the GWs can be produced and could be observed by the future experiments for the GW detection~\cite{Schwaller:2015tja}.

In this paper, we study the spectra of GWs due to the dynamical chiral symmetry breaking (D$\chi$SB) in the dQCD Eq.\,\eqref{original Lag} with dark pion as a dark matter.
We assume that the symmetry breaking pattern $\text{SU}\fn{N_f}_L \times \text{SU}\fn{N_f}_R \to \text{SU}\fn{N_f}_V$ takes place.
For the WIMP dark matter scenario, we have to specify the connection between the dark sector and the SM sector in order to have a process 2 DM\,$\to$\, 2 SM.
In this work, we introduce the dark sector Eq.\,\eqref{original Lag} as a classically scale invariant extension of the SM with a singlet scalar field $S$~\cite{Hur:2007uz,Hur:2011sv,Heikinheimo:2013fta,Farzinnia:2013pga,Foot:2013hna,Holthausen:2013ota,Kubo:2014ida,Ametani:2015jla} since this model is a simple model for the dark pion as a WIMP dark matter. 
The dark quark mass $m_{ij}$ is given by the Yukawa interaction $y_{ij}S\bar \psi_i\psi_j$.
Due to the strong dynamics of dQCD, the chiral and scale symmetries are dynamically broken and then the electroweak (EW) scale are generated.
The dark pion is produced via the D$\chi$SB and obtains the mass being proportional to $y_{ij}\bra S\ket$.
In contrast, for the SIMP dark matter scenario, the annihilation process of dark pions is given by the dynamics of dark pions only.
Therefore, we do not have to specify the connection to the SM.\footnote{Nevertheless, the dark sector has to be connected to the SM sector in order to thermalize DM.
The kinetic interaction between hidden pions and SM particles has been discussed in Ref.\,\cite{Kamada:2016ois}.
See also \cite{Bernal:2015ova}.
}
Introducing chiral perturbation theory, we describe the pion dynamics and evaluate the relic abundance. 

To evaluate the spectra of GWs, we need to investigate the effective potential and the chiral phase transition at finite temperature. 
However, the treatment of the strong dynamics described by Eq.\,\eqref{original Lag} is complicated. 
Therefore, we use the effective theory approach to the dark sector: $\Lag_\text{dQCD}\simeq \Lag_\text{eff}$.
In this work, we introduce the linear sigma model and investigate the chiral phase transition at finite temperature for both WIMP and SIMP dark matter scenarios.
It has been discussed in Refs.~\cite{Lenaghan:2000ey,Roder:2003uz} that the linear sigma model actually describes the critical phenomena of the chiral phase transition well.
Since the critical temperature and order of the phase transition are important for spectra of GWs, it would be useful to extract the essence of the phase transition from the linear sigma model.

This paper is organized as follow:
In the next section, we discuss the dark pion as the WIMP within the classically scale invariant extension of the SM.
The allowed parameter space which can explain the observed values of DM relic abundance is investigated.
At a benchmark point, we calculate the spectra of GWs background.
In section~\ref{simp case}, the SIMP dark matter is discussed.
We show the allowed region for the pion mass and decay constant. 
The constraints come from the observed relic abundance of DM and pion-pion scattering amplitude.
A possibility of producing the GWs is discussed within our setups. 
We summarize our work in section~\ref{summary}.
In Appendix, several supplementary formulas are given.

\section{Weakly interacting massive particle}\label{wimpsection}
In this section, we consider the dark pion as a WIMP dark matter.
In order to generate annihilation process of dark pions into the SM particles, we need to fix a connection between the dark and the SM sector.
To this end, we here consider the following classically scale invariant Lagrangian~\cite{Hur:2007uz,Hur:2011sv},
\al{
\Lag_\text{total}=\Lag_\text{SM}|_{m_\text{H}^2\to 0}+\Lag_\text{dark},
}
where $\Lag_\text{SM}$ is the SM Lagrangian and
\al{
\Lag_\text{dark}=-\frac{1}{4}F_{\mu\nu}^aF^a{}^{\mu\nu} + \bar \psi_i (\delta _{ij}i\Slash D - y_{ij}S) \psi_j
+\frac{1}{2}(\p_\mu S)^2 -\frac{\lambda_S}{4}S^4 +\frac{\lambda_{HS}}{2}S^2H^\dagger H
\label{darksectorLag}
}
is the Lagrangian of the dark sector.
Here, $S$ is a singlet real scalar field and is coupled to the SM Higgs doublet field $H$ via the Higgs portal coupling $\lambda_{HS}$.
The dark quark masses are generated by the Yukawa interaction.
Hereafter, we assume that $\text{SU}\fn{N_f}_V$ symmetry is not broken, that is, the Yukawa coupling matrix is proportional to the identity matrix: $y_{ij}=y\delta_{ij}$.
Due to the strong dynamics of dQCD, the D$\chi$SB takes place, and the dark quark obtains the dynamical mass.
Its mass scale can be the origin of the EW scale through the mediator $S$ and the Higgs portal coupling, i.e. the Higgs mass term is given by $m_\text{H}^2 \sim \lambda_{HS}\bra S\ket^2$.
If $\lambda_{HS}$ is of order $10^{-3}$, the scale of the dark sector is TeV order.
After the D$\chi$SB, the stable dark pions appear thanks to the flavor symmetry.
These dark pions have a finite mass via the Yukawa interaction being the bare dark quark mass $m_q=y\bra S\ket$.

The literature~\cite{Holthausen:2013ota} has studied the model~Eq.\,\eqref{darksectorLag} using the Nambu--Jona-Lasinio model~\cite{Nambu:1961tp,Nambu:1961fr,Hatsuda:1994pi} for the dQCD sector. 
The relic abundance of the dark pion and the chiral and EW phase transitions at finite temperature have been investigated.
Here, we use the linear sigma model and follow the formulations given in \cite{Holthausen:2013ota} for the dark pion as a candidate of dark matter.

\subsection{Vacuum and mass spectrum}
To investigate the vacuum of Eq.\,\eqref{darksectorLag}, we employ the linear sigma model for the dQCD sector.
In this work, the $N_f=3$ case is considered since the Columbia plot indicates that the chiral phase transition in three-flavor QCD becomes first-order for the small bare quark masses. 
Its basic formulations are shown in Appendix~\ref{explicit linier sigma model}.
Using Eq.\,\eqref{tree level potential}, the effective potential at the tree level for Eq.\,\eqref{darksectorLag} is approximately given by
\al{
V_\text{tree}\fn{{\bar \sigma},h,S}&= 
\frac{m^2}{2} {\bar \sigma}^2
-\frac{c}{3\sqrt{6}}{\bar\sigma}^3 
+\frac{1}{4}\left( \lambda_1+\frac{\lambda_2}{3}\right) {\bar\sigma}^4
-j{\bar\sigma}
+\frac{\lambda_S}{4}S^4 -\frac{\lambda_{HS}}{4}S^2h^2
+\frac{\lambda_H}{4}h^4,
} 
where $\bar \sigma$ and $h$ are the scalar meson field, and the physical mode of the Higgs field is parametrized as $H=(0,h/\sqrt{2})^T$, respectively.
The expectation value $\bra \bar \sigma \ket$ corresponds to the chiral condensate.
Since we have $j=m^2m_q$ and $m_q=yS$ (see Eq.\,\eqref{relation3}), we see that the explicit chiral symmetry breaking term is proportional to $S$, namely, $j=m^2yS$.
Note that the higher loop effects qualitatively do not change the results~\cite{Lenaghan:1999si,Lenaghan:2000ey}.

Let us here set the parameters.
Since the linear sigma model is employed for dQCD sector, there are no constraints on the parameters in this sector from experiments.
In this work, we use similar values given in \cite{Lenaghan:1999si}, where the pion mass spectra in real QCD are determined in the linear sigma model, namely,
\al{
m^2&=(810)^2\, \text{GeV}^2,&
c&=5600\, \text{GeV},&
\lambda_1&=-15,&
\lambda_2&=50.&
\label{darksector parameters}
}
For the coupling constants for the scalar sector, the following parameters are taken as an example for illustration:
\al{
y&=0.2293,&
\lambda_S&=0.2,&
\lambda_{HS}&=0.01,&
\lambda_H&=0.132.&
\label{darksector higgs parameters}
}
We will scan these parameters to be satisfied the constraints from the Higgs mass, the EW vacuum and the relic abundance of the DM.
In this case, the vacuum is determined as
\al{
\bra \bar\sigma \ket &=1079.27\,\text{GeV},&
\bra h \ket &=181.61\,\text{GeV},&
\bra S \ket &=933.14\,\text{GeV}.&
}
Since the EW vacuum is $v_h=246$\,GeV, we can define the rescale factor $\zeta=246\, \text{GeV}/\bra h \ket\simeq 1.35454$.
Using this factor, the dimensionful parameters are rescaled and then we obtain
\al{
\bar \sigma_0&=\bra \bar\sigma \ket =1461.91\,\text{GeV},&
v_h&=\bra h \ket =246\,\text{GeV},&
v_S&=\bra S \ket =1263.97\,\text{GeV}.&
\label{vacua}
}
Note that the dimensionful parameters $m^2$ and $c$ are also rescaled by $\zeta$. 

We next calculate the masses for $\sigma$, $h$, $S$ and $\pi$.
The two point functions at the tree level for the CP-even scalar fields are 
\al{
\Gamma_{\sigma\sigma}\fn{p^2}&=p^2-m^2+\sqrt{\frac{2}{3}}c\,{\bar \sigma}-(3\lambda_1+\lambda_2){\bar \sigma}^2,&
\Gamma_{\sigma h}\fn{p^2}&=\Gamma_{h\sigma}\fn{p^2}=0,&\nn
\Gamma_{\sigma S}\fn{p^2}&=\Gamma_{S\sigma }\fn{p^2}=\frac{ym^2}{2},&
\Gamma_{hh}\fn{p^2}&=p^2 -3\lambda_H \bra h\ket ^2 +\frac{\lambda_{HS}}{2}\bra S\ket^2,&\nn 
\Gamma_{Sh}\fn{p^2}&=\Gamma_{hS}\fn{p^2}=\frac{\lambda_{HS}}{2}\bra h\ket \bra S\ket,&
\Gamma_{SS}\fn{p^2}&=p^2 -3\lambda_S \bra S\ket^2 +\frac{\lambda_{HS}}{2}\bra h\ket^2,&
}
Then the masses for $\sigma$, $h$, $S$ are given by the poles of the two-point function matrix
\al{
\Gamma\fn{p^2}=\pmat{
\Gamma_{hh}\fn{p^2} & \Gamma_{Sh}\fn{p^2} & \Gamma_{\sigma h}\fn{p^2} \\
\Gamma_{Sh}\fn{p^2} & \Gamma_{SS}\fn{p^2} & \Gamma_{\sigma S}\fn{p^2}  \\
\Gamma_{\sigma h}\fn{p^2} &  \Gamma_{\sigma S}\fn{p^2} & \Gamma_{\sigma\sigma}\fn{p^2}
},
}
i.e, the physical masses satisfy 
\al{
\Gamma_{ij}\fn{m_k^2}\xi_j^{(k)}=0.
}
We obtain the masses and the corresponding eigenvectors
\al{
m_1&=126.387\,\text{GeV},&
(\xi_1^{(1)},\xi_2^{(1)},\xi_3^{(1)})&=(0.999999, 0.00164989, 6.7083\times10^{-11}),
\label{higgsmass}
\\
m_2&=978.914\,\text{GeV},&
(\xi_1^{(2)},\xi_2^{(2)},\xi_3^{(2)})&=(-0.00164989, 0.999999, 6.10691\times10^{-8}),
\label{sSmass}
\\
m_3&=1683.88\,\text{GeV},&
(\xi_1^{(3)},\xi_2^{(3)},\xi_3^{(3)})&=(3.36741\times 10^{-11}, -6.10691\times 10^{-8}, 1).
\label{sigmamass}
}
The relation between the flavor eigenstates $(h,S,\sigma)$ and the mass eigenstates $(\phi_1,\phi_2,\phi_3)$ is given by
\al{
\pmat{
h\\S\\ \sigma
}
=\pmat{
\xi_1^{(1)} & \xi_1^{(2)} & \xi_1^{(3)}\\
\xi_2^{(1)} & \xi_2^{(2)} & \xi_2^{(3)}\\
\xi_3^{(1)} & \xi_3^{(2)} & \xi_3^{(3)}
}
\pmat{
\phi_1\\ \phi_2 \\ \phi_3
}.
}  
Hereafter we write the mass eigenvalues as $m_1=m_h$, $m_2=m_S$ and $m_3=m_\sigma$.

Since the pion field is CP-odd and is not mixed with the other fields, its mass at the tree level (see Eq.\,\eqref{Pmass}) is
\al{
m_\pi=\sqrt{m^2 
-\frac{c}{\sqrt{6}}\bra\bar \sigma\ket
+\left( \lambda_1 +\frac{\lambda_2}{3}\right)\bra\bar \sigma\ket^2}
\simeq 488.497\, \text{GeV}.
}

\subsection{Dark pion relic abundance}
\begin{figure}
\begin{center}
\includegraphics[width=150mm]{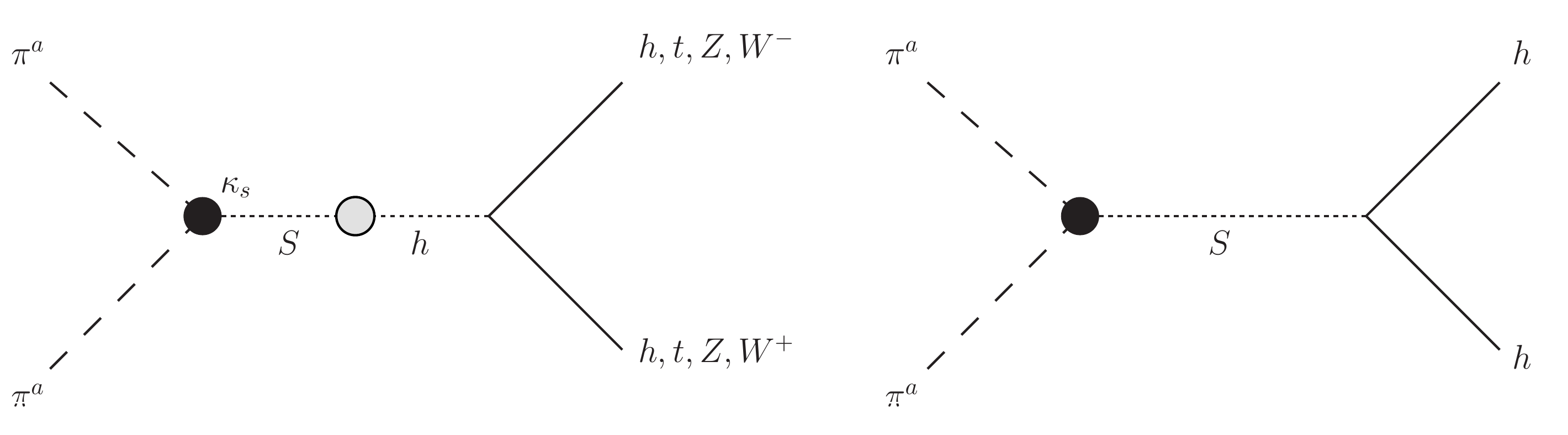}
\end{center}
\caption{Annihilation processes of the dark pions into the SM particles. The black circle is the effective vertex of $\pi^2S$}
\label{s channel diagrams}
\end{figure}
We here calculate the thermal relic abundance of the WIMP DM.
The annihilation processes of two dark pions at the tree level are shown in Fig.~\ref{s channel diagrams}.
The thermally averaged cross section for their annihilation process is given by
\al{
\bra \sigma v\ket
&=\frac{1}{32\pi m_\pi^3N_\pi^2}\bigg[ (m_\pi^2-M_W^2)^{1/2}a_W+(m_\pi^2-M_Z^2)^{1/2}a_Z\nn
&\quad +(m_\pi^2-M_t^2)^{1/2}a_t +(m_\pi^2-m_h^2)^{1/2}a_h\bigg] + {\mathcal O}\fn{v^2},
}
where $N_\pi$ is the number of pion,\footnote{
In the present analysis with a finite anomaly term $c\neq 0$ we have $N_\pi=N_f^2-1$.
Note that we have $N_\pi=N_f^2$ with the vanishing anomaly $c=0$ since the dark pion and the $\eta'$-meson become degenerate in mass: $(m_P^2)_{00}=(m_P)_{ii}^2=:m_\pi^2$ for $i=1,...,8$; see a discussion below Eq.\,\eqref{etapionmass}.} 
$v$ is the DM relative velocity; and $M_W\simeq 80.4\,\text{GeV}$, $M_Z\simeq 91.2\,\text{GeV}$ and $M_t\simeq 173\,\text{GeV}$ are the $W$-boson, the $Z$-boson and the top-quark masses, respectively.
We also define the coefficients as
\al{
a_W&=16\left(\frac{\kappa_s}{v_h}\right)^2 |\Delta_{hS}|^2m_\pi^4\left(1 -\frac{M_W^2}{m_\pi^2} + \frac{3}{4}\frac{M_W^4}{m_\pi^4}\right),\\
a_Z&=8\left(\frac{\kappa_s}{v_h}\right)^2 |\Delta_{hS}|^2m_\pi^4 \left(1-\frac{M_Z^2}{m_\pi^2} +\frac{3}{4}\frac{M_Z^4}{m_\pi^4} \right),\\
a_t&=24\left( \frac{\kappa_s}{v_h}\right)^2|\Delta_{hS}|^2M_t^2 m_\pi^2 \left(1-\frac{M_t^2}{m_\pi^2}\right),\\
a_h&=8\left( \frac{\kappa_s}{v_h}\right)^2 \left| 3\lambda_H v_h\Delta_{hS}-\frac{\lambda_{HS}v_S}{2}\Delta_{SS}\right|^2,
}
and the propagators as
\al{
\Delta_{hS}&=\frac{\xi_2^{(2)}\xi_1^{(2)}}{4m_\pi^2-m_S^2+i\gamma_S m_S}+\frac{\xi_2^{(1)}\xi_1^{(1)}}{4m_\pi^2-m_h^2},
\label{propagatorhs}
\\
\Delta_{SS}&=\frac{\xi_2^{(2)}\xi_2^{(2)}}{4m_\pi^2-m_S^2+i\gamma_S m_S}+\frac{\xi_2^{(1)}\xi_2^{(1)}}{4m_\pi^2-m_h^2},
\label{propagatorss}
}
with the decay width of $S$,
\al{
\gamma_S=\frac{(\lambda_{HS}v_S)^2}{8\pi m_S^2}\sqrt{\frac{m_S^2}{4}-m_h^2}.
}
Since the decay through the Higgs mixing is suppressed by $v_h/\bar{\sigma}_0$ and $v_h/v_S$, 
this formula is good approximation for the decay of $S$. 
\begin{figure}
\begin{center}
\includegraphics[width=150mm]{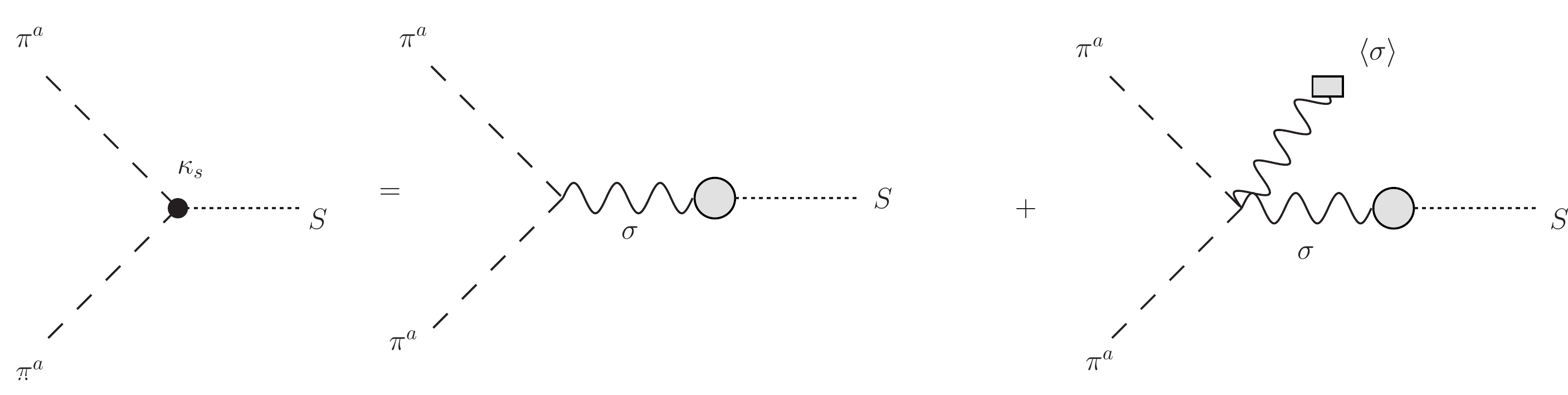}
\end{center}
\caption{The $s$-channel effective vertex of $\pi^2S$. The gray circle denotes the coupling constant ${\tilde y}$}
\label{s channel vertex}
\end{figure}
$\kappa_s$ is the coupling constant of the $s$-channel effective interaction $\pi^2S$ whose diagrams are shown in Fig.~\ref{s channel vertex}.
From the 6th term in Eq.\,\eqref{lagrangianbroken} and the explicit breaking term it is calculated as
\al{
\kappa_s=2\left[\sqrt{\frac{3}{2}}c
+3\left(\lambda_1+\frac{\lambda_2}{3}\right)\bra \bar\sigma \ket
\right]\frac{m^2y}{4m_\pi^2-m_\sigma^2}.
}
\label{kappas}

We now show the formula for the dark matter relic abundance~\cite{Griest:1988ma},
\al{
\Omega_{\text{dark}\,\pi }{\hat h}^2 =\frac{Y_\infty s_0 m_\pi}{\rho_c/{\hat h}^2},
\label{relic}
}
where $s_0=2970/\text{cm}^3$ is the entropy density of the universe at present and $\rho_c/{\hat h}^2=1.05\times 10^{-5}\, \text{GeV}/\text{cm}^3$ is the critical density divided by the square of the dimensionless Hubble parameter.
$Y_\infty$ is the solution at $x=m_\pi/T\to\infty$ of the Boltzmann equation 
\al{
\frac{\df Y}{\df x}=-0.264g_*\fn{T}^{1/2}\left(\frac{M_\text{Pl}m_\pi}{x^2}\right)\bra \sigma v\ket (Y^2-\bar Y^2),
\label{boltzmanneq}
}
with the Planck mass $M_\text{Pl}=1.22\times 10^{19}\, \text{GeV}$ and the degrees of freedom of relativistic particles $g_*\fn{T}\simeq 115.75$ at TeV scale temperature.
Note that the degrees of freedom of the dark pions and the singlet-scalar are included in $g_*\fn{T}$.
$\bar Y$ is $Y$ in thermal equilibrium;
\al{
{\bar Y}=\frac{45 x^2N_\pi}{4\pi^4g_*}K_2\fn{x},
\label{themaltequiY}
}
with the modified Bessel function of the second kind $K_2\fn{x}$.
The approximated solution of Eq.\,\eqref{boltzmanneq} is given by
\al{
Y_\infty^{-1}=0.264g_*^{1/2}\frac{M_\text{Pl}m_\pi}{N_\pi}\frac{\bra \sigma v\ket }{x_f},
}
where $x_f$ is the ratio $m_\pi/T$ at the freeze-out temperature and is obtained from~\cite{Griest:1988ma}
\al{
x_f=\log \fn{\frac{0.095M_\text{Pl}m_\pi\bra  \sigma v\ket }{(g_*x_f)^{1/2}}}.
}
When we use the parameter set Eq.\,\eqref{darksector parameters} and Eq.\,\eqref{darksector higgs parameters}, we obtain
\al{ 
\Omega_{\text{dark}\, \pi}{\hat h}^2\simeq 0.1168.
\label{cosmologicalvalues}
}

\begin{figure}
\begin{center}
\includegraphics[width=60mm]{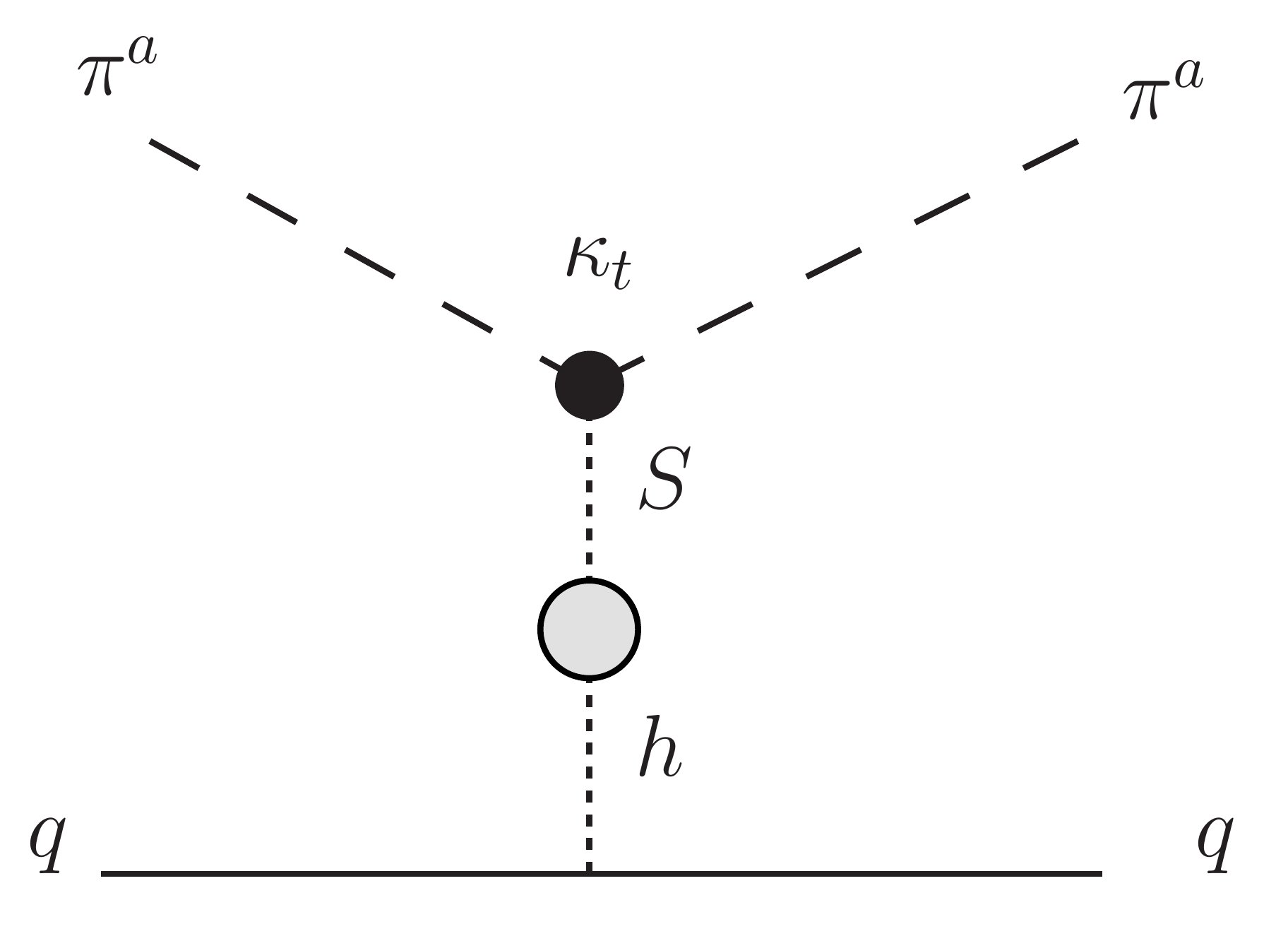}
\end{center}
\caption{The scattering process between the dark pion and the quarks in the nucleon. }
\label{cross section off}
\end{figure}
\begin{figure}
\begin{center}
\includegraphics[width=160mm]{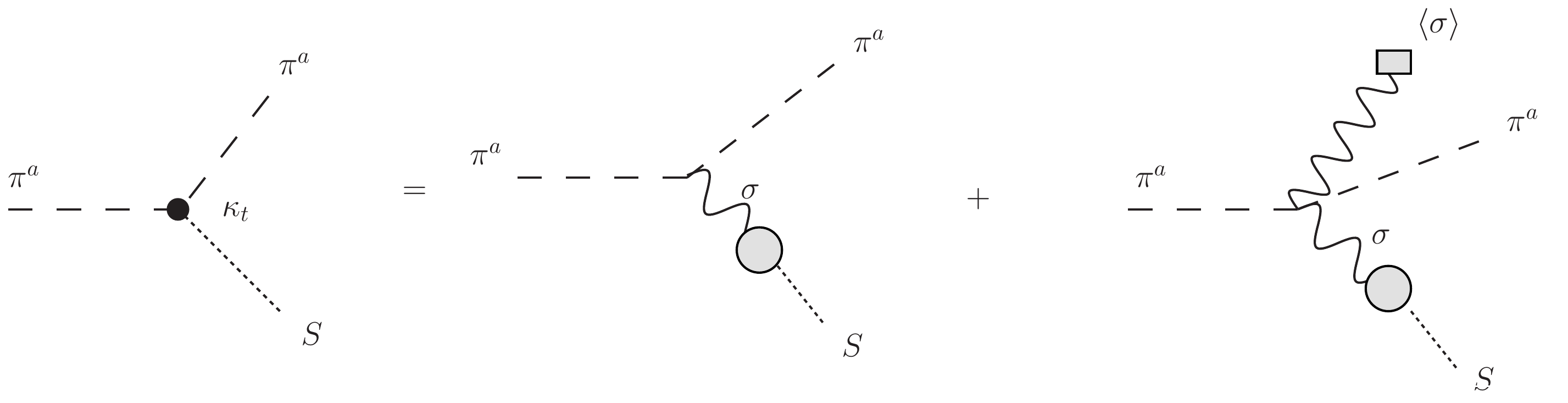}
\end{center}
\caption{The $t$-channel effective vertex of $\pi^2S$.}
\label{t channel vertex}
\end{figure}
Moreover, we give the spin-independent elastic cross section off the nucleon~\cite{Barbieri:2006dq} as shown in Fig.~\ref{cross section off}, 
\al{
\sigma_\text{SI}=\frac{1}{4\pi}\left(\frac{\kappa_t }{v_h}\right)^2\left[ {\hat f}m_N\left( \frac{\xi_2^{(2)}\xi_1^{(2)}}{m_S^2} +\frac{\xi_2^{(1)}\xi_1^{(1)}}{m_h^2}\right) \right]^2\left( \frac{m_N}{m_N+m_\pi}\right)^2,
}
where $m_N$ is the nucleon mass and ${\hat f}\sim 0.3$ is the nucleonic matrix element.\footnote{
$\hat f$ is evaluated as
\al{
\hat f=\sum_{q=u,...,t}f_q^N=\frac{2}{9}+\frac{7}{9}(f_u^N+f_d^N+f_s^N)= 0.305\pm 0.009,
\nonumber
}
where we used values of  $f_q^N$ given in \cite{Hoferichter:2015dsa}.
The quark contents of the nucleon $f_q^N$ are calculated by lattice simulation~\cite{Oksuzian:2012rzb,Junnarkar:2013ac}, dispersion relations~\cite{Hoferichter:2015dsa} and chiral effective field theories~\cite{Alarcon:2011zs,Alarcon:2012nr}.}
Here, $\kappa_t$ is the $t$-channel effective vertex which is shown in Fig.~\ref{t channel vertex} and is evaluated as
\al{
\kappa_t=2\left[\sqrt{\frac{3}{2}}c
+3\left(\lambda_1+\frac{\lambda_2}{3}\right)\bra \bar\sigma \ket
\right]\frac{m^2y}{-m_\sigma^2}.
}
\label{kappat}
For the parameter set Eq.\,\eqref{darksector parameters} and Eq.\,\eqref{darksector higgs parameters}, we have
\al{
\sigma_\text{SI}\simeq 1.0658\times 10^{-49}\,\text{cm}^2.
}

\begin{figure}
\begin{center}
\includegraphics[width=8cm]{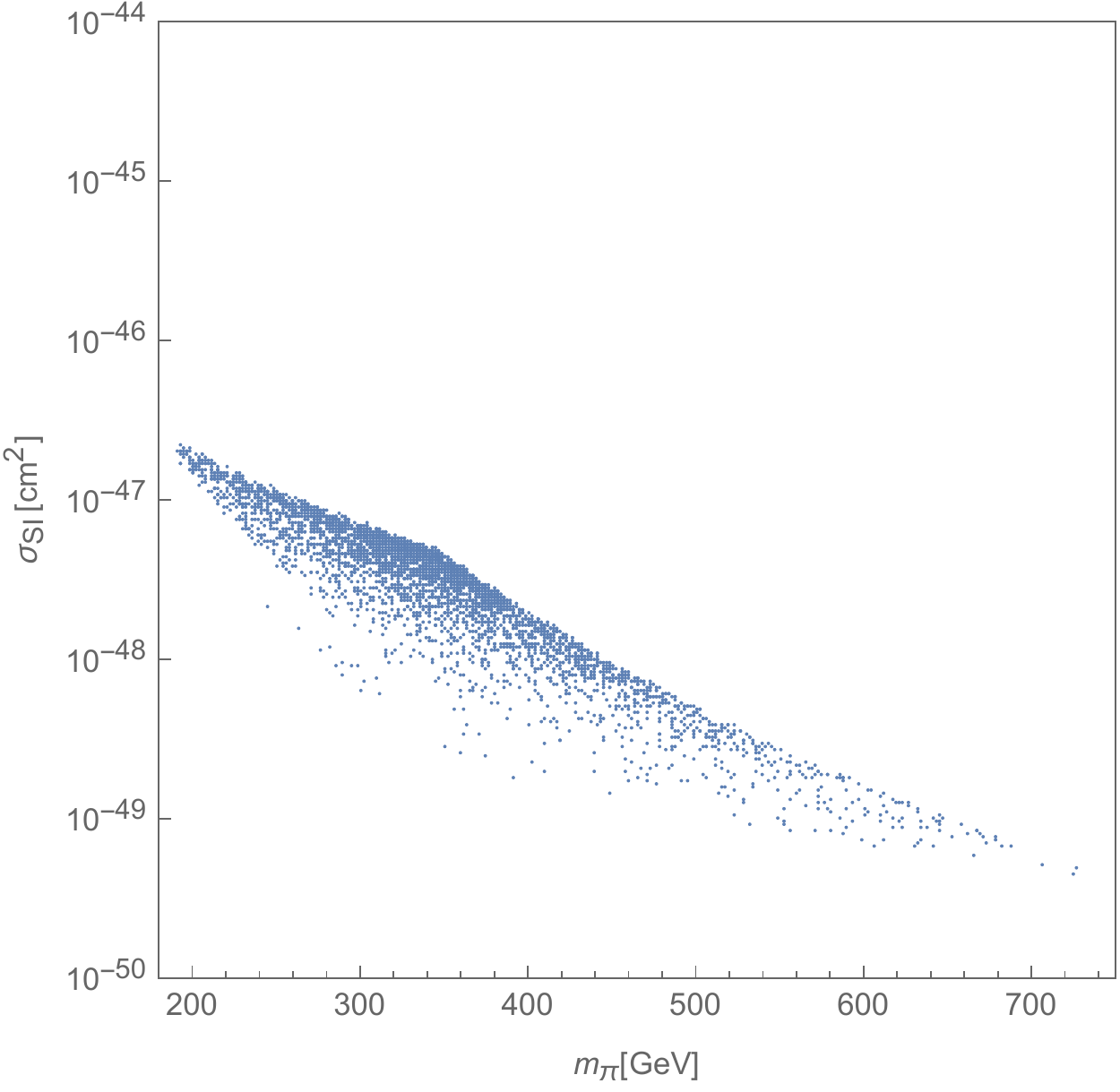}
\includegraphics[width=8cm]{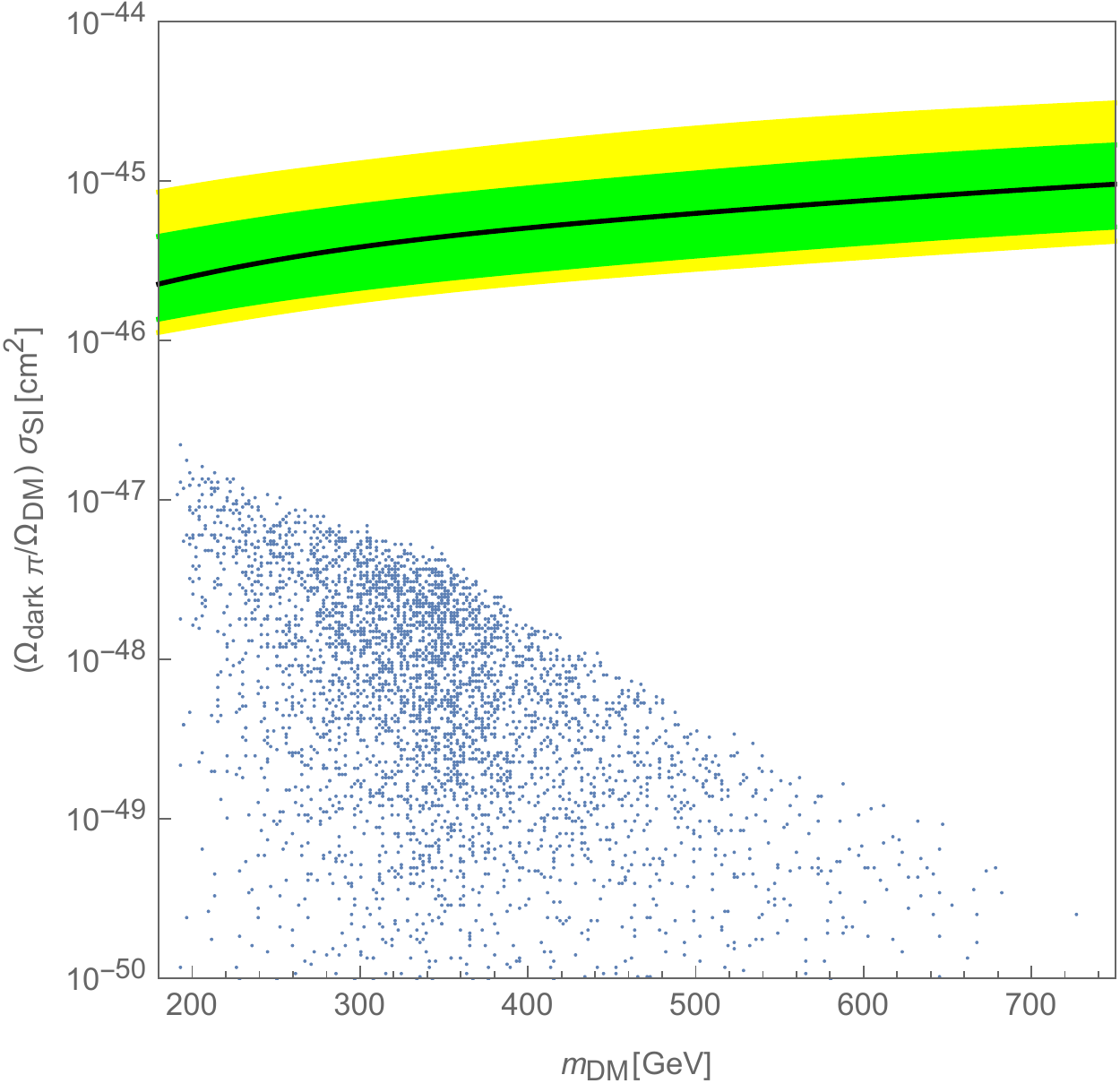}
\caption{Left: The spin-independent elastic cross section off the nucleon $\sigma_\text{SI}$ as a function of the dark pion mass $m_\pi$.
Right: $\sigma_\text{SI}$ rescaled by $\Omega_{\text{dark}\,\pi}/\Omega_\text{DM}$.
The blue dots show the allowed region in the model. The black solid line denotes the central value of XENON1T with one (green) and two (yellow) $\sigma$ bands~\cite{Aprile:2017iyp}.
}
\label{SigmaSImpi}
\end{center}
\end{figure}

We now search the parameter region where we have $\Omega_{\text{dark}\,\pi} {\hat h}^2\lsim \Omega_\text{DM} {\hat h}^2\simeq 0.12$~\cite{Ade:2015xua}, $m_h=126\,\text{GeV}$~\cite{Chatrchyan:2012xdj,Aad:2012tfa} and $v_h=246\,\text{GeV}$.
Here it is imposed that the relic abundance of dark pion is smaller than observed value of DM since the dark $\eta'$ meson in the isospin limit and the dark baryon may as well be stable.
Indeed, in real QCD the proton is stable.
For the parameters of the dark sector, we use Eq.\,\eqref{darksector parameters} and vary $y$, $\lambda_S$, $\lambda_{HS}$ and $\lambda_H$.
The $\sigma_\text{SI}$--$m_\pi$ plot is shown in Fig.~\ref{SigmaSImpi}.\footnote{
We have searched the allowed region in the parameter space $\lambda_H\in (0.131,0.132)$, $\lambda_{HS} \in (0.01,0,05)$, $\lambda_{S}\in (0.01,0.5)$ and $y \in (0.1,1.0)$.
}
The left panel of Fig.~\ref{SigmaSImpi} shows the spin-independent elastic cross section off the nucleon $\sigma_\text{SI}$ as a function of the dark pion mass $m_\pi$.
Since we impose that the relic abundance of the dark pion is less than $\Omega_\text{DM} {\hat h}^2\simeq 0.12$, we show the rescaled spin-independent elastic cross section by the ratio $\Omega_{\text{dark}\,\pi}/\Omega_\text{DM}$ and the latest result of XENON1T experiment~\cite{Aprile:2017iyp} in the right panel.
The rescaled cross section $(\Omega_{\text{dark}\,\pi}/\Omega_\text{DM})\sigma_\text{SI}$ is below the upper bound (${\mathcal O}\fn{10^{-46}}\,\text{cm}^2$) of the latest result of XENON1T experiment~\cite{Aprile:2017iyp}.
XENON1T~\cite{Aprile:2012zx,Aprile:2015uzo,Aprile:2017iyp} has a sensitivity ${\mathcal O}\fn{10^{-47}}\,\text{cm}^2$ in future.
Therefore, the light dark pion could be tested by XENON1T in the future.

Note that within our effective theory approach, the effective coupling constant $\kappa_s$ cannot be large to explain the relic abundance $\Omega_{\text{dark}\,\pi}{\hat h}^2\lsim 0.12$.
This fact is reported in \cite{Holthausen:2013ota} where the NJL model is employed as well.
Hence, we use the resonant effect, namely, $2m_\text{DM}\simeq m_S$ at which the propagators Eq.\,\eqref{propagatorhs} and Eq.\,\eqref{propagatorss} are enhanced.
This constraint could be relaxed by imposing the $\text{U}\fn{1}$ charge on the dark quarks or breaking $\text{SU}\fn{3}_V$ flavor symmetry group to smaller one with the non-degenerate Yukawa coupling matrix $y_{ij}$.
This possibilities actually are shown in the NJL model approach~\cite{Kubo:2014ida,Ametani:2015jla}.

\subsection{Gravitational waves from chiral phase transition for WIMP case}
\begin{figure}
\begin{center}
\includegraphics[width=80mm]{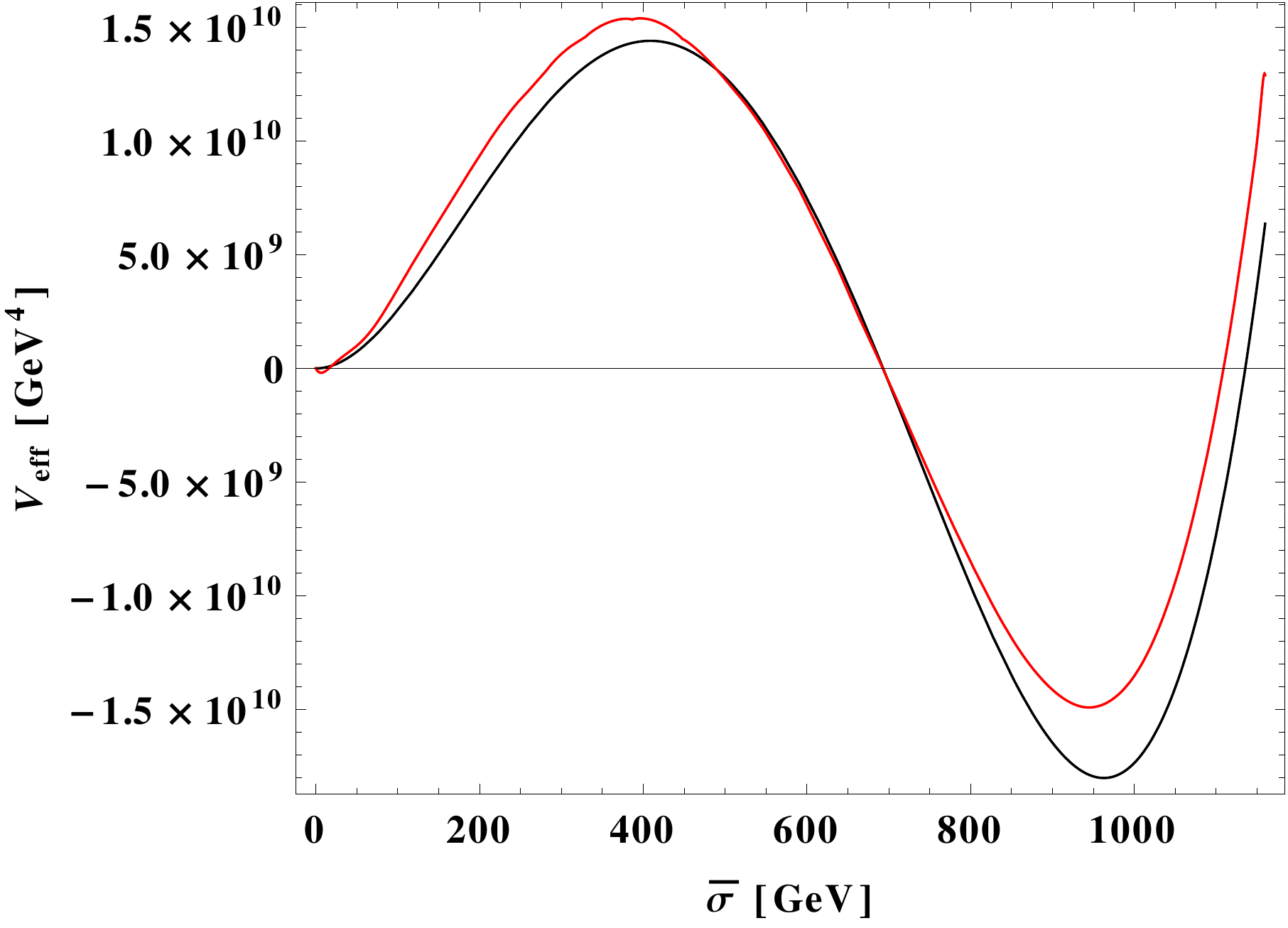}
\includegraphics[width=80mm]{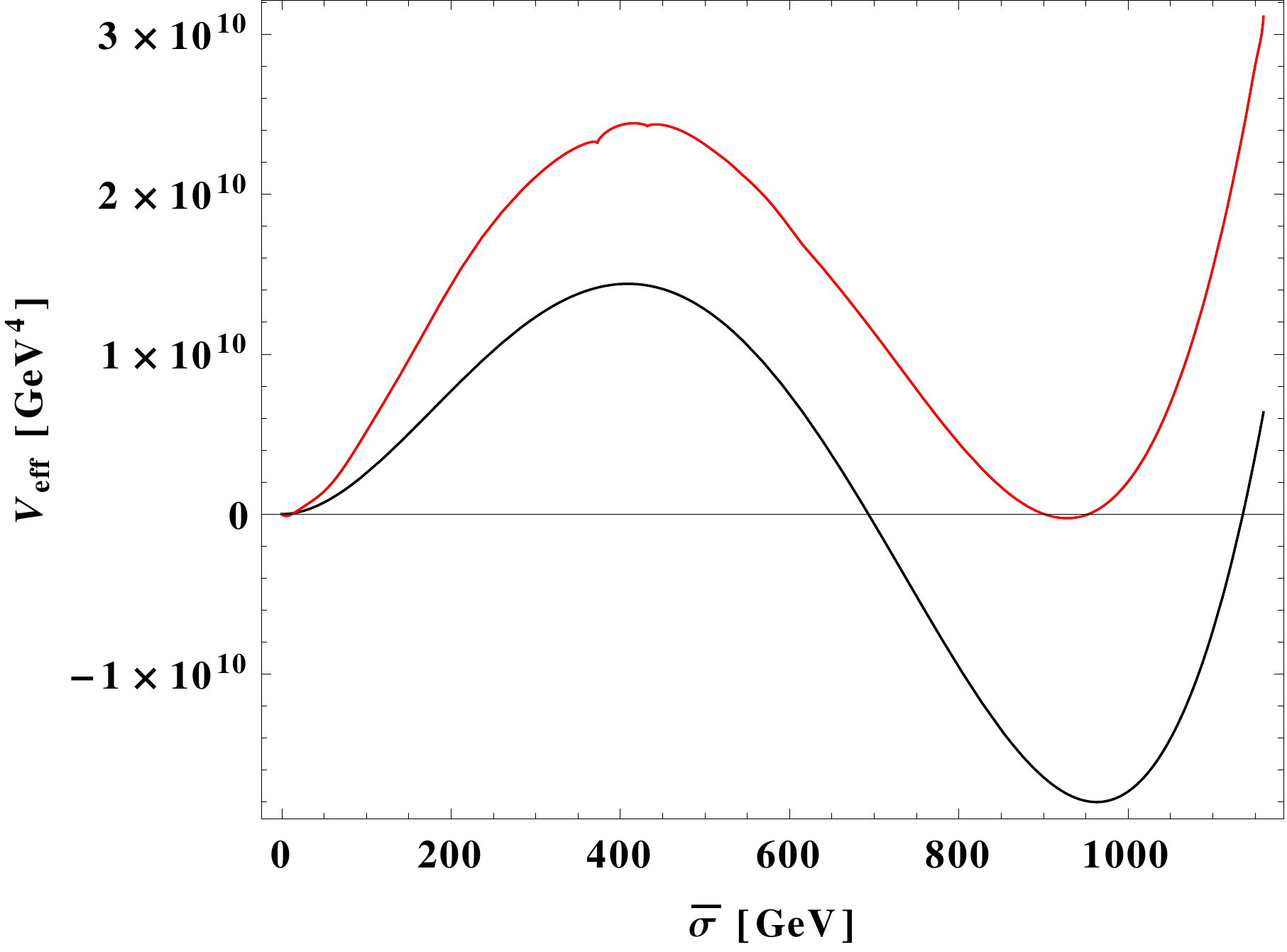}
\includegraphics[width=100mm]{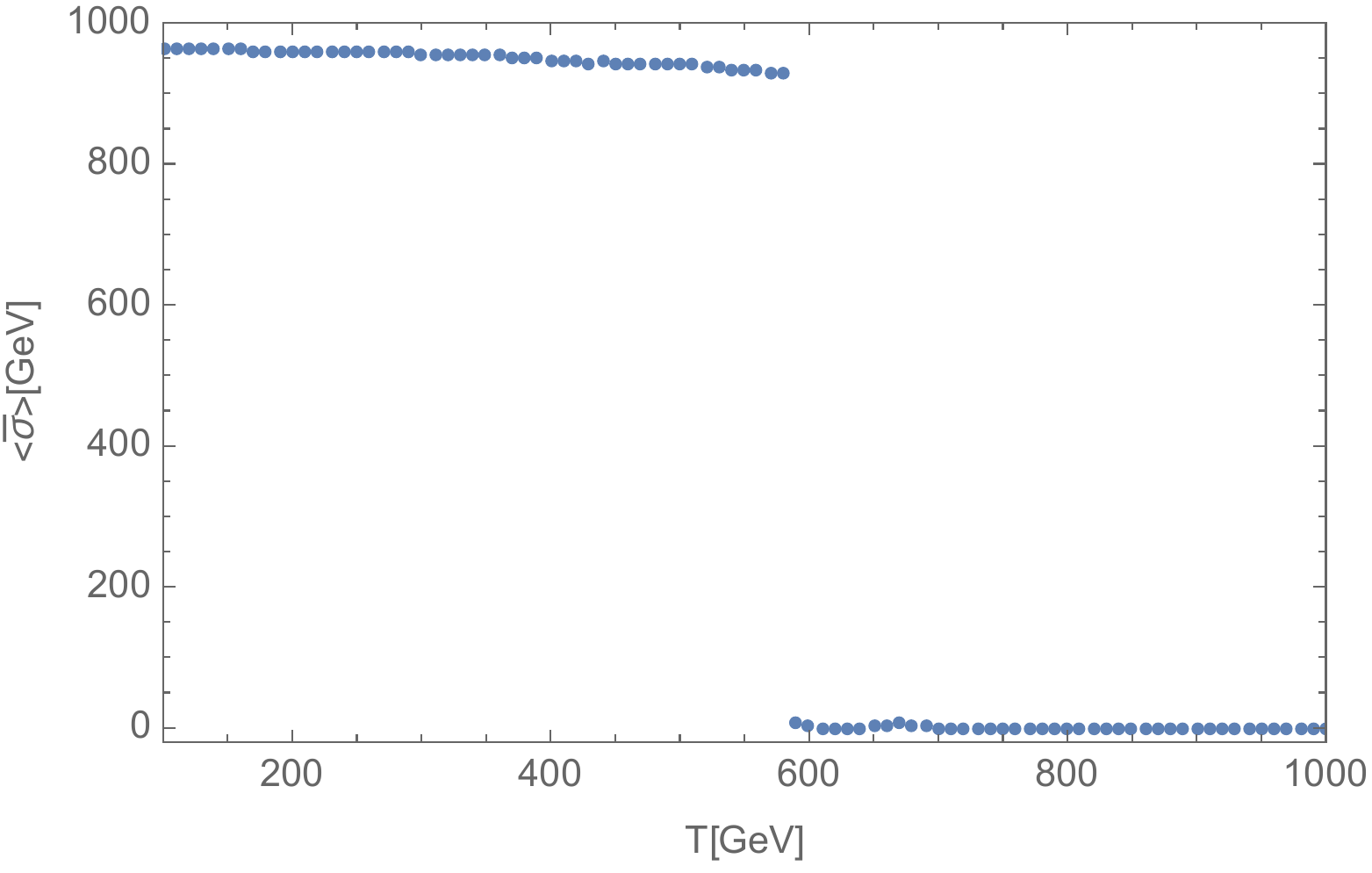}
\end{center}
\caption{The black and red lines are the effective potential at zero and finite temperatures, respectively.
Top-Left: The effective potential at the cosmological phase transition temperature $T_t\simeq 430\,\text{GeV}$. Top-Right: The effective potential at the critical temperature $T_c\simeq 690\,\text{GeV}$ at which the symmetric and the broken vacua degenerate.
Bottom: The behavior of the expectation value $\bra \bar \sigma \ket$ for varying temperature.
}
\label{potentials}
\end{figure}
\begin{figure}
\begin{center}
\includegraphics[width=80mm]{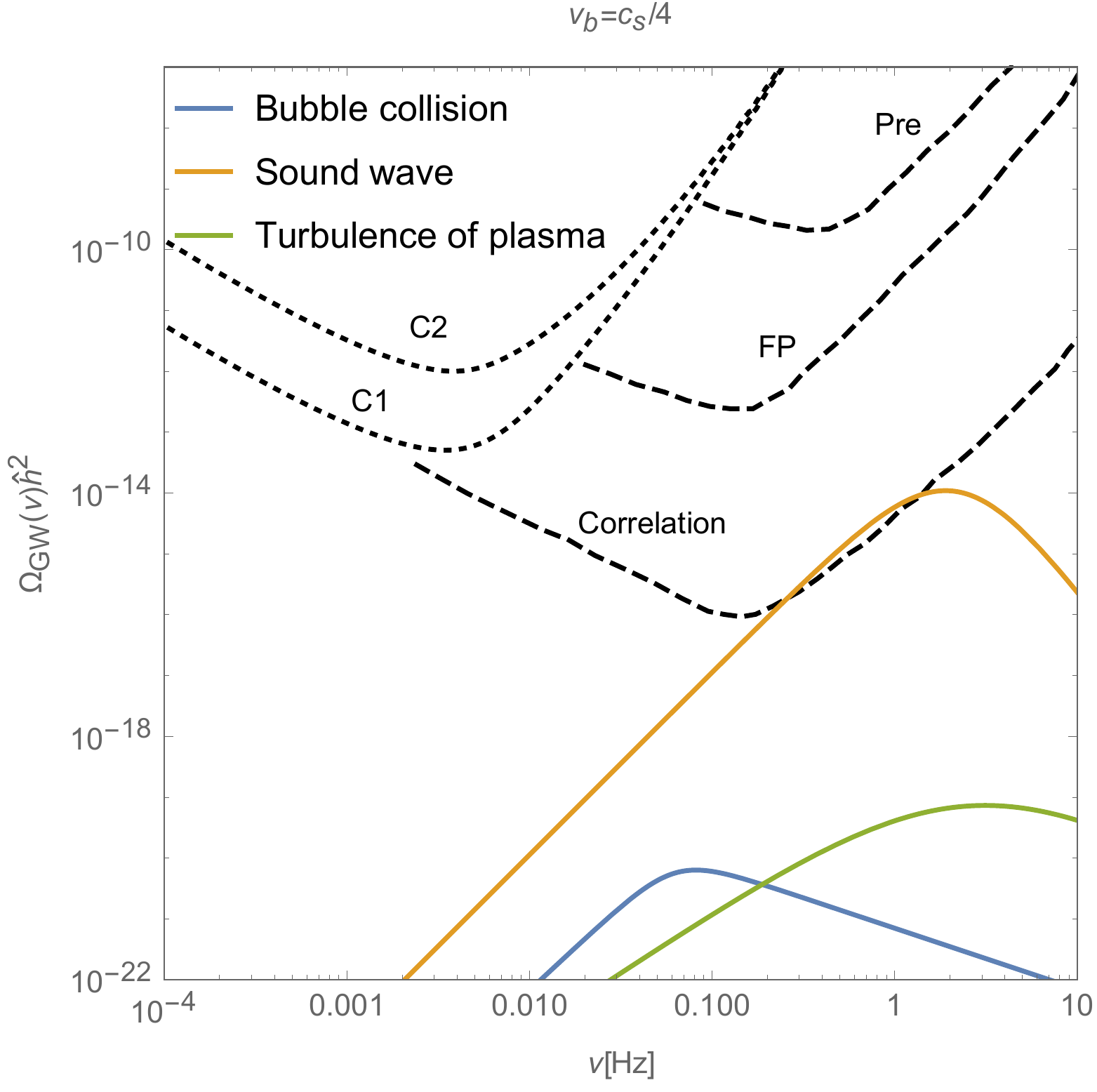}
\includegraphics[width=80mm]{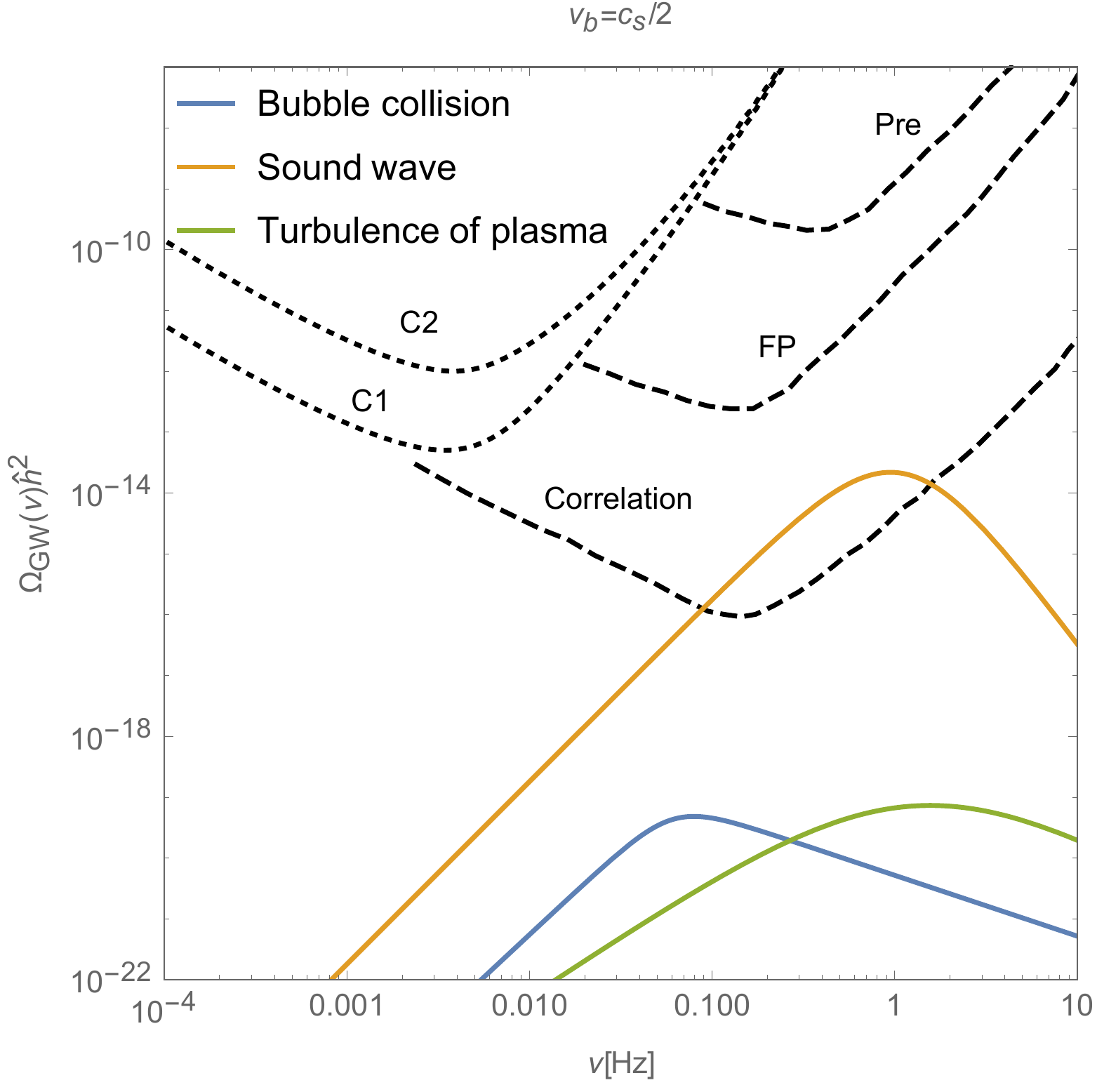}
\includegraphics[width=80mm]{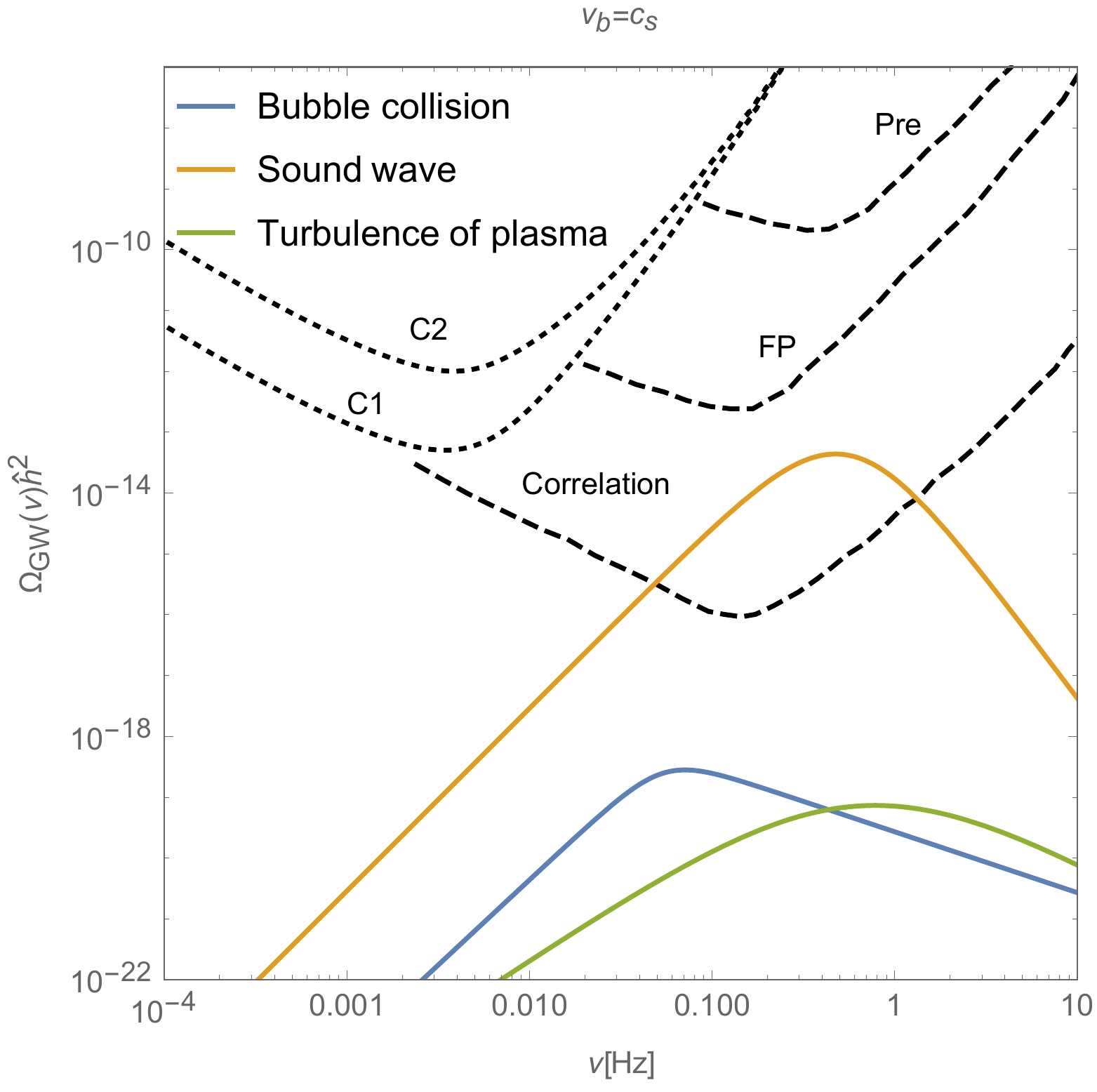}
\includegraphics[width=80mm]{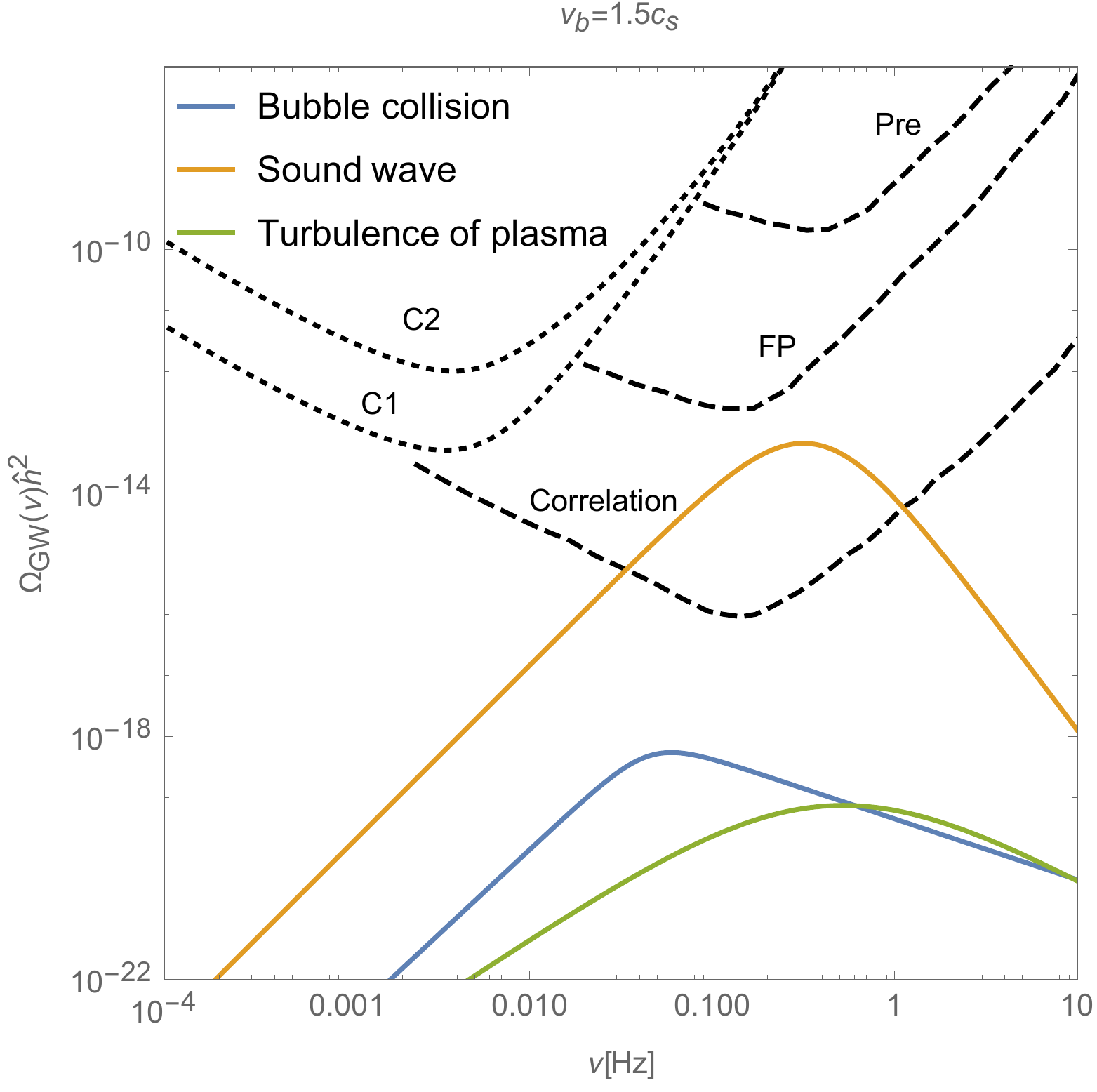}
\caption{The spectra of GWs. The dotted curves show the sensitivities of eLISA~\cite{Seoane:2013qna,Audley:2017drz}. The labels (``C1" and ``C2") denotes the different configurations listed in table.~1 in \cite{Caprini:2015zlo}. Their data sets are taken from \cite{eLisadetasets}.
The dashed lines represent different designs of DECIGO~\cite{Seto:2001qf,Kawamura:2006up,Kawamura:2011zz}.
}
\label{wimpgw}
\end{center}
\end{figure}
We investigate the GW background spectra from the chiral phase transition in the classically scale invariant model Eq.\,\eqref{darksectorLag}.
To this end, using the Cornwall-Jackiw-Tomboulis (CJT) formalism~\cite{Cornwall:1974vz} discussed in appendix~\ref{EFPlinear}, we evaluate the effective potential of the linear sigma model at finite temperature.
We use the parameter set Eq.\,\eqref{darksector parameters} as the benchmark point which yield the physical values Eq.\,\eqref{vacua}, Eq.\,\eqref{higgsmass}--Eq.\,\eqref{sigmamass} and Eq.\,\eqref{cosmologicalvalues}.
Since $\bra h \ket < \bra S\ket < \bra \bar \sigma \ket$, we can focus on only the dark sector dynamics, i.e., consider $V_\text{eff}\fn{\bar \sigma,h=0,S=0}$ at chiral phase transition temperature.
The effective potential in this case is shown in Fig.~\ref{potentials}.

As shown in Appendix~\ref{FGWsformula}, the GW background spectra are produced by three mechanisms, namely, the bubble collision $\Omega_\text{coll}{\hat h}^2$, the sound wave $\Omega_\text{sw}{\hat h}^2$ and the turbulence of plasma $\Omega_\text{MHD}{\hat h}^2$.
These GW background spectra are characterized by three model-dependent parameters.
One is ``cosmological" phase transition temperature $T_t$ at which the Euclidean action~Eq.\,\eqref{eucideanaction} satisfies the criterion of the phase transition in the expanding universe Eq.\,\eqref{phcriterion}.
Other two parameters are the latent heat $\alpha$ and the (inverse) duration time $\tilde \beta$ which are given in Eq.\,\eqref{alphaparameter} and Eq.\,\eqref{tildebetaparameter}, respectively.

In the top side of Fig.~\ref{potentials}, we show the effective potential at zero temperature (black line), the cosmological phase transition temperature $T_t\simeq 430\,\text{GeV}$ (red line in the left-hand side panel) and  the critical temperature $T_c\simeq 690\,\text{GeV}$ (red line in the right-hand side panel) at which the vacua degenerate.
The $T$ dependence of the expectation value is shown in the bottom side of Fig.~\ref{potentials}.
We find that the strong first-order phase transition takes place and the cosmological phase transition takes place at $T_t\simeq 430\,\text{GeV}$.
At the cosmological phase transition temperature, we obtain\footnote{
Note that the value of $\alpha$ corresponds to $\Delta\epsilon\fn{T_t}/T_t^4\simeq 1.33$.
This value is comparable to $\Delta\epsilon\fn{T}/T^4=0.75\pm 0.17$ in \cite{Shirogane:2016zbf} where the lattice simulation for the pure Yang-Mills theory has been performed.
}
\al{
\alpha&=0.038,&
\tilde \beta&=3169.25.&
\label{latenduration}
}
Besides, the GW background spectra depend on the velocity of bubble walls $v_b$ and the number of degrees of freedom for relativistic particles $g_*^t=g_*\fn{T_t}$.
As dynamics of the phase transition, there are three possible cases: ``non-runaway bubbles"; ``runway bubbles in plasma"; and ``runaway bubbles in vacuum"~\cite{Caprini:2015zlo}.
In the first case, bubbles in a plasma reach a terminal velocity being smaller than the speed of light, and the energy of scalar field can be negligible, i.e., $\Omega_\text{GW}\fn{\nu} {\hat h}^2\simeq \left[\Omega_\text{SW}\fn{\nu}  + \Omega_\text{MHD}\fn{\nu} \right] {\hat h}^2$.
The spectra of the bubble collision become smaller than the others.
In the second case, three mechanisms contribute to the total spectra of GWs.
In the last case, the bubble velocity quickly reaches the speed of light ($v_b=1$) and the spectrum from the bubble collision becomes dominant.
See \cite{Caprini:2015zlo} for detailed discussion on the bubble dynamics.
Although the velocity of bubble wall should be determined by dynamics of bubbles, it cannot be precisely evaluated by the analyses and then is treated as a free parameter.
In this work, we assume ``non-runaway" bubbles expanding with velocities near the speed of sound $c_s=0.577$.\footnote{
When the vacuum energy released at the cosmological phase transition is adequately large, the velocity of bubble wall reaches to the speed of light ($v_b=1$).
We expect that this situation would be somewhat extreme.
}

At $T=T_t\simeq 430\,\text{GeV}$, we have $g_*^t\simeq 115.75$.
The formulas for the GW spectra are given in Appendix~\ref{FGWsformula}.
Using the values $T_t\simeq 430\,\text{GeV}$ and Eq.\,\eqref{latenduration} with several bubble wall velocities; $v_b=c_s/4$, $c_s/2$, $c_s$ and $1.5c_s$, the GW background spectra from three sources are shown in Fig.~\ref{wimpgw}, where the peaks of GW background spectra and frequencies become
\begin{itemize}
\item for $v_b=c_s/4$,
\al{
{\tilde \Omega}_\text{coll}{\hat h}^2&\simeq 6.31\times 10^{-21},&
\tilde \nu_\text{coll}&\simeq 0.081651\,\text{Hz},&\\
{\tilde \Omega}_\text{SW}{\hat h}^2&\simeq 1.09\times 10^{-14},&
\tilde \nu_\text{SW}&\simeq 1.8990\,\text{Hz},&\\
\frac{{\tilde \Omega}_\text{MHD}{\hat h}^2}{\left( 1+8\pi {\tilde \nu}_\text{MHD}/h_t \right)}&\simeq 9.11\times 10^{-19},&
\tilde \nu_\text{MHD}&\simeq 2.6119\,\text{Hz};&
}
\item for $v_b=c_s/2$,
\al{
{\tilde \Omega}_\text{coll}{\hat h}^2&\simeq 4.81\times 10^{-20},&
\tilde \nu_\text{coll}&\simeq 0.079538\,\text{Hz},&\\
{\tilde \Omega}_\text{SW}{\hat h}^2&\simeq 2.1756\times 10^{-14},&
\tilde \nu_\text{SW}&\simeq 0.94952\,\text{Hz},&\\
\frac{{\tilde \Omega}_\text{MHD}{\hat h}^2}{\left( 1+8\pi {\tilde \nu}_\text{MHD}/h_t \right)}&\simeq 9.11\times 10^{-19},&
\tilde \nu_\text{MHD}&\simeq 1.3060\,\text{Hz};&
}
\item for $v_b=c_s$,
\al{
{\tilde \Omega}_\text{coll}{\hat h}^2&\simeq 2.79\times 10^{-19},&
\tilde \nu_\text{coll}&\simeq 0.071073\,\text{Hz},&\\
{\tilde \Omega}_\text{SW}{\hat h}^2&\simeq 4.35\times 10^{-14},&
\tilde \nu_\text{SW}&\simeq 0.47476\,\text{Hz},&\\
\frac{{\tilde \Omega}_\text{MHD}{\hat h}^2}{\left( 1+8\pi {\tilde \nu}_\text{MHD}/h_t \right)}&\simeq 9.11\times 10^{-19},&
\tilde \nu_\text{MHD}&\simeq 0.65298\,\text{Hz};&
}
\item for $v_b=1.5c_s$,
\al{
{\tilde \Omega}_\text{coll}{\hat h}^2&\simeq 5.40\times 10^{-19},&
\tilde \nu_\text{coll}&\simeq 0.059895\,\text{Hz},&\\
{\tilde \Omega}_\text{SW}{\hat h}^2&\simeq 6.53\times10^{-14},&
\tilde \nu_\text{SW}&\simeq 0.31651\,\text{Hz},&\\
\frac{{\tilde \Omega}_\text{MHD}{\hat h}^2}{\left( 1+8\pi {\tilde \nu}_\text{MHD}/h_t \right)}&\simeq 9.11\times 10^{-19},&
\tilde \nu_\text{MHD}&\simeq 0.43532\,\text{Hz}.&
}
\end{itemize}
Note that for instance, ${\tilde \Omega}_\text{MHD}{\hat h}^2\simeq 1.885 \times 10^{-13}>{\tilde \Omega}_\text{coll}{\hat h}^2$ in the $v_b=c_s$ case.
In Fig.~\ref{wimpgw}, we also plot the sensitivities of eLISA~\cite{Seoane:2013qna} with the dotted lines and DECIGO~\cite{Seto:2001qf,Kawamura:2006up,Kawamura:2011zz} with the dashed lines.
As mentioned above, the spectrum from the bubble collision is weaker than the others in case of non-runaway bubbles.
Although the turbulence of plasma produces strong spectrum of GWs, its peak is suppressed by due to the dependence on the factor $(1+8\pi {\tilde \nu}_\text{MHD}/h_t)$ with the Hubble rate.
Consequently, the spectrum from the sound wave becomes dominant.
We see that the GW background spectrum from the sound wave could be observed by DECIGO.

\section{Strongly interacting massive particle}\label{simp case}
It has been shown in \cite{Hochberg:2014kqa} that the dark pion can be a candidate of SIMP DM.
The dynamics of the pion in the low energy region can by described by chiral perturbation theory.
The WZW term, which has five-point pion interactions, plays a crucial role as the $3\to 2$ annihilation process of pion.
Since this annihilation process does not involve the SM particles, we here do not specify a connection to the SM sector.

In this section, we discuss the dynamics of the dark pion as the SIMP DM by using chiral perturbation theory.
In order for the dark pion to phenomenologically be relevant, the following constraints have to be satisfied:
From the bullet cluster~\cite{Markevitch:2003at}, the $2\to 2$ process (pion-pion scattering) cross section is constrained as $\sigma_{2\to 2}/m_\pi \lsim 1\, \text{cm}^2/\text{g}$; the dark pion mass must not become larger than the perturbative limit, i.e., $m_\pi/f_\pi \lsim 4\pi$.
Ref.~\cite{Hansen:2015yaa} has pointed out that taking the contributions from the higher order in chiral perturbation theory into account, the dark pion as the SIMP is strongly constrained. 

Following Refs.~\cite{Hochberg:2014kqa,Hansen:2015yaa}, we start with setting the formulations of chiral perturbation theory and the cross sections of the pion scattering processes for a general flavor number.
Then, for $N_f=3$ we investigate the allowed region where the dark pion satisfies the constraints.

\subsection{Chiral perturbation theory}
Let us consider the D$\chi$SB pattern $G=\SU\fn{N_f}_L\times \SU\fn{N_f}_R \to H=\SU\fn{N_f}_V$.
We parametrize the Goldstone boson manifold $G/H$ as
\al{
u&=\exp\fn{\frac{i}{\sqrt{2}f_\pi}\pi},&
\pi&=\pi^aX_a,&
\label{GB parametrize}
}
where $f_\pi$ is the dark pion decay constant,\footnote{
In real QCD, $f_\pi\simeq 93$\, MeV.
}
and $X^a$ are the broken generators normalized as $\tr (X^aX^b)=\delta^{ab}$.
Under the group transformation the quantity $u$ transforms as
\al{
u\to g_R u h^\dagger=h ug_L^\dagger,
}
where $g_{L(R)}$ and $h$ are a group element of $\SU\fn{N_f}_{L(R)}$ and the subgroup $H$, respectively.
We here introduce the following quantities:
\al{
u_\mu&=i(u^\dagger \p_\mu u-u\p_\mu u^\dagger),
\label{umu}
\\
\chi_\pm&=\chi u^\dagger \pm u\chi^\dagger.
\label{chipm}
}
The quantity $\chi$ is given as $\chi=2B M$ where $B$ is a value being proportional to the chiral condensate, and $M$ is a bare quark mass matrix.

In {chiral perturbation theory} the chiral Lagrangian is expanded into the polynomials of the dark pion mass and external momenta of the dark pion, that is,
\al{
\Lag_\text{ChPT}=\Lag_\text{LO} +\Lag_\text{NLO} + {\mathcal O}\fn{p^6}.
}
Using the quantities Eq.\,\eqref{umu} and Eq.\,\eqref{chipm}, each term can be written down as follow.
The first term is the lowest order (LO), i.e. ${\mathcal O}\fn{p^2}$ and becomes
\al{
\Lag_\text{LO} =\frac{f_\pi^2}{4}\tr[ u_\mu u^\mu +\chi_+ ],
}
where ``$\tr$" denotes the trace for the flavor space.

The next leading order (NLO) Lagrangian with ${\mathcal O}\fn{p^4}$~\cite{Gasser:1984gg} is given as
\al{
\Lag_\text{NLO} &=L_0\tr[ u_\mu u_\nu u^\mu u^\nu] 
+L_1\tr[ u_\mu u^\mu ] \tr[ u_\nu u^\nu]
+L_2\tr[ u_\mu u_\nu ] \tr[ u^\mu u^\nu]
+L_3\tr[ u_\mu u^\mu u_\nu u^\nu] \nn
&\quad
+L_4\tr[ u^\mu u_\mu] \tr[ \chi_+]
+L_5\tr[ u^\mu u_\mu \chi_+]
+L_6\tr[ \chi_+]^2
+L_7\tr[ \chi_-] ^2
+\frac{1}{2}L_8\tr[ \chi_+^2+\chi_-^2],
}
where only the terms contributing to {pion-pion scattering} have been kept.
Since the NLO Lagrangian contains the loop corrections, the coefficients $L_i$ have the ultraviolet divergences, and then they have to be renormalized.
Using the dimensional regularization and the $\overline{\text{MS}}$ scheme, we have
\al{
L_i=L_i^r-\frac{\Gamma_i}{32\pi^2}\left(\frac{2}{\epsilon} +\log\fn{4\pi}-\gamma_E+1 \right),
}
where $\epsilon=4-d$, $\gamma_E\simeq 0.577$ is the {Euler}-Mascheroni constant, and the coefficients $\Gamma_i$ are given in~\cite{Gasser:1983yg,Bijnens:2009qm}.

The results for the $2\to 2$ scattering in both the NLO and the NNLO are in the uncertainty band.
This work does not focus on precise calculations, and then we take the higher order contributions up to the NLO into account for the $2\to 2$ scattering.

\subsection{Pion-pion scattering process}
The cross section for {pion-pion scattering} is given by
\al{
\sigma_{2\to 2}=\frac{|T|^2}{128\pi^2N_\pi^2m_\pi^2},
}
where $N_\pi=N_f^2-1$ is the number of the broken generators, and the amplitude $T$ for {pion-pion scattering} process is
\al{
T^{{abcd}}\fn{s,t,u}=\bra \pi^c\fn{p_3}\pi^d\fn{p_4}| \pi^a\fn{p_1}\pi^b\fn{p_2}\ket.
}
We introduce the dimensionless Mandelstam variables,
\al{
s&=\frac{(p_1+p_2)^2}{m_\pi^2},&
t&=\frac{(p_1-p_3)^2}{m_\pi^2},&
u&=\frac{(p_1-p_4)^2}{m_\pi^2},&
\label{Mandelstam variables stu}
}
which satisfy $s+t+u=4$.
The amplitude $T$ is
\al{
T^{{abcd}}\fn{s,t,u}&=\xi^{abcd}B\fn{s,t,u}+\xi^{acdb}B\fn{t,u,s}+\xi^{adbc}B\fn{u,s,t}\nn
&\quad+\delta^{ab}\delta^{cd}C\fn{s,t,u} + \delta^{ac}\delta^{bd}C\fn{t,u,s}+\delta^{ad}\delta^{bc}C\fn{u,s,t},
}
where $\xi^{abcd}=\tr[X^aX^bX^cX^d] + \tr[X^aX^dX^cX^b]$, especially, for the $\SU\fn{N_f}_L\times \SU\fn{N_f}_R \to \SU\fn{N_f}_V$ case, we have
\al{
\xi^{abcd}=\frac{2}{3}(\delta^{ab}\delta^{cd}-\delta^{ac}\delta^{bd}+\delta^{ad}\delta^{bc})
+d^{abe}d^{cde}-d^{ace}d^{bde}+d^{ade}d^{bce}.
}
The functions $B$ and $C$ in {chiral perturbation theory} are expanded as
\al{
B\fn{s,t,u}&=B_\text{LO}\fn{s,t,u}+B_\text{NLO}\fn{s,t,u},\\
C\fn{s,t,u}&=C_\text{LO}\fn{s,t,u}+C_\text{NLO}\fn{s,t,u}.
}
These functions are written by the dimensionless Mandelstam variables, the flavor number $N_f$ and the renormalized coefficients $L_i^r$.
Their explicit forms are shown in Appendix~\ref{BC functions}.
Note that in this work we use ``$p^4$ fit" data given in table~1 of \cite{Bijnens:2014lea} for the values of $L_i^r$.

\subsection{$3\to 2$ pion annihilation process}
We next give a formulation in order to evaluate the relic abundance of the dark pion.
The number density of the dark pion is reduced by the $3\to 2$ annihilation process which can be described by the Wess-Zumino-Witten term~\cite{Wess:1971yu,Witten:1983tw,Witten:1983tx},
\al{
\Lag_\text{WZW}= \frac{N_c}{240\pi^2}\int_0^1 \df\alpha\int \df^4x\, \epsilon^{ABCDE}\tr[u_A^\alpha u_B^\alpha u_C^\alpha u_D^\alpha u_E^\alpha],
}
where we defined
\al{
u^\alpha&=\exp\fn{\frac{i\alpha}{\sqrt{2}f_\pi}X^a\phi^a},&
u^\alpha_A&=i(u^\alpha{}^\dagger\p_A u^\alpha-u\p_A u^\alpha{}^\dagger).&
}

The full Lagrangian is given by
\al{
\Lag_\text{full}= \Lag_\text{ChPT} + \Lag_\text{WZW}.
}
More explicitly, using the pion field $\pi$ given in Eq.\,\eqref{GB parametrize} the Lagrangian is
\al{
\Lag_\text{full}
&=\frac{1}{2}\tr\,[\p_\mu \pi \p^\mu \pi] 
-\frac{m_\pi^2}{2}\tr\, [\pi^2]
+\frac{m_\pi^2}{12f_\pi^2}\tr\,[\pi^4]
-\frac{1}{6f_\pi^2}\tr[\pi^2\p^\mu \pi \p_\mu \pi -\pi \p^\mu \pi\p_\mu \pi]
\nn
&\quad +\frac{2N_c}{15\pi^2f_\pi^5}\epsilon^{\mu\nu\rho\sigma}\text{tr}[\pi \p_\mu \pi \p_\nu \pi \p_\rho \pi \p_\sigma \pi]+{\mathcal O}\fn{\pi^6}.
}

The thermally averaged cross section at NLO is calculated as~\cite{Hochberg:2014kqa}
\al{
\bra \sigma v^2\ket_{3\to2}^\text{NLO}=\frac{5\sqrt{5}}{2048\pi^5x^2}\frac{N_c^2m_\pi^5}{f_\pi^{10}}\frac{t^2}{N_\pi^3},
}
where 
\al{
t^2=\frac{1}{5!}\sum T^2_{\{ijklm\}}.
}
In the breaking pattern $\SU\fn{N_f}_L\times \SU\fn{N_f}_{R}\to \SU\fn{N_f}_V$ case, we have
\al{
t^2&=\frac{4}{3}N_f(N_f^2-1)(N_f^2-4).
} 

When the NNLO effects are taken into account, the cross section becomes~\cite{Hansen:2015yaa}
\al{
\bra \sigma v^2\ket_{3\to2}^\text{NNLO}=\bra \sigma v^2\ket_{3\to2}^\text{NLO}\left( 1+\frac{m_\pi^2}{f_\pi^2}(a_wL+b_w)\right),
}
where we have introduced the shorthand notations
\al{
L&=\frac{1}{16\pi^2}\log\fn{\frac{m_\pi^2}{\mu^2}},
}
with the renormalization point $\mu^2$, and the coefficients are given by
\al{
a_w&=-\frac{77}{6}\simeq -12.83,&
b_w&=\frac{5\sqrt{5}}{288\pi^2}\log\fn{\frac{9+\sqrt{45}}{9-\sqrt{45}}}-\frac{7}{96\pi^2}
\simeq 1.83\times 10^{-4}.&
}

\subsection{Dark pion relic abundance}
The Boltzmann equation for the $3\to2$ process is given by~\cite{Hochberg:2014kqa}
\al{
\frac{\df Y}{\df x} = -0.116g_*\fn{T}^{1/2}\left(\frac{M_\text{Pl}m_\pi^4}{x^5}\right)(Y^3-Y^2{\bar Y})\bra \sigma v^2\ket_{3\to2},
\label{boltz}
}
where $x=m_\pi/T$ and $\bar Y$ is given in Eq.\,\eqref{themaltequiY}.
The $2\to 2$ annihilation process ($2\, \pi\to $ 2\,SM particles) is neglected since it is subdominant process~\cite{Hochberg:2014kqa}.
Using the solution $Y_\infty$ of the Boltzmann equation Eq.\,\eqref{boltz} at $x\to \infty$, one can obtain the relic abundance which is defined in Eq.\,\eqref{relic}.

In the next subsection, we show the region of $m_\pi$ and $f_\pi$, which satisfy the dark matter relic abundance, the constraints for {pion-pion scattering} and the perturbative limit of {chiral perturbation theory}.

\subsection{Numerical analysis}
\begin{figure}
\begin{center}
\includegraphics[width=100mm]{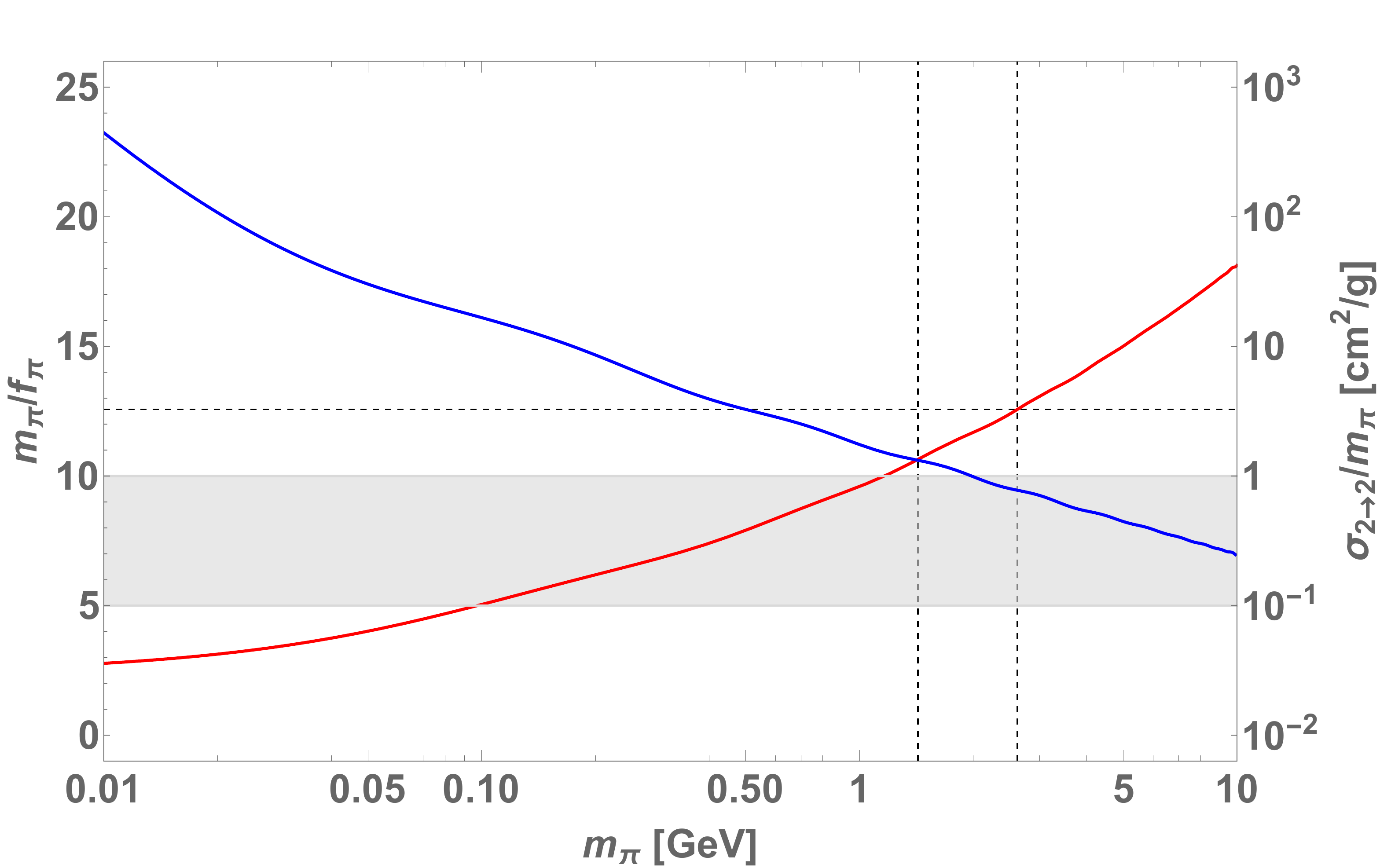}
\end{center}
\caption{
The red line denotes the values of $m_\pi$ and $m_\pi/f_\pi$ which satisfy the dark matter relic abundance $\Omega_\text{DM}{\hat h}\simeq 0.12$.
The blue line shows the dependence of {pion-pion scattering} cross section $\sigma_{2\to 2}/m_\pi$ on $m_\pi$. 
The constraint for $\sigma_{2\to 2}/m_\pi$ on $m_\pi$ with the uncertainty is shown by the gray band.
The dotted horizontal line stands for the perturbative limit for {chiral perturbation theory}.
The region surrounded by the dotted vertical lines is allowed.
}
\label{allowed region}
\end{figure}
We have set up the formulations to evaluate {pion-pion scattering} cross section and the relic density of the dark pion.
When evaluating them numerically, we take into account the NNLO contribution for the thermal average cross section and set the variables as
\al{
N_c&=16,&
N_f&=3,&
s&=4,&
t&=u=0,&
\mu^2&= 4m_\pi^2,
}
and $g_*\fn{T}\simeq g_s\fn{T}\simeq 100$.
Since the Columbia plot indicates that the chiral phase transition in three-flavor QCD become first-order for the small bare quark masses, $N_f=3$ case is considered.
We show the allowed region of the dark pion mass and their decay constant in Fig.~\ref{allowed region}.
The dark matter relic abundance $\Omega_\text{DM}{\hat h}\simeq 0.12$ is satisfied on the red line.
The blue line stands for the dependence of {pion-pion scattering} cross section $\sigma_{2\to 2}/m_\pi$ on $m_\pi$.
The pion-pion scattering cross section should satisfy the limit $\sigma_{2\to 2}/m_\pi \lsim 0.1\text{--}1\,\text{cm}^2/\text{g}$ which comes from the evidence of the gravitational lensing of the bullet cluster~\cite{Markevitch:2003at} and the simulations for the halo cluster~\cite{Zavala:2012us,Rocha:2012jg}  .
The allowed parameter space from the uncertainty is shown by the gray band.
The blue line has to be below the line $\sigma_{2\to 2}/m_\pi=1\,\text{cm}^2/\text{g}$.
The dotted horizontal line stands for the perturbative limit for {chiral perturbation theory}.
The red and blue lines have to be below this limit.
Therefore, the mass and the decay constant of the dark pion have to satisfy the region surrounded by the dotted vertical lines.

\subsection{Can gravitational wave be produced from chiral phase transition in SIMP case?}
We next consider chiral phase transition in the case with the dark pion as the SIMP dark matter.
Actually, we will find the chiral phase transition generally becomes crossover,
 and then, there would be no GW signal within the present setup.

Since in the allowed region the dark pion mass,
 which is given by $m_\pi^2=j/\bra \bar \sigma\ket$,
 is $\mathcal O\fn{1}\, \text{GeV}$,
 the coupling of the symmetry breaking term $j$ should be large as $\mathcal O\fn{1}\, \text{GeV}^3$.
This fact induces that the chiral phase transition tends to be crossover as one can see from the Columbia plot~\cite{Brown:1990ev}.
Although $m_\pi$ can be larger for the smaller $\bra \bar \sigma\ket$, this case is limited by the perturbative limit $m_\pi/f_\pi<4\pi$ since it is proportional to $\bra \bar \sigma\ket^{-3/2}$

To check that the chiral phase transition numerically becomes crossover,
 let us rewrite the potential in terms of $m_\pi$ and $f_\pi$,
 which satisfy the allowed region given in the previous subsection.
From the relation between $m_\pi$ and $f_\pi$, the parameters in the effective potential,
\al{
U\fn{\bar \sigma}&=\frac{m^2}{2}{\bar \sigma}^2-\frac{c}{3\sqrt{6}}{\bar \sigma}^3+\frac{\lambda}{4}{\bar \sigma}^4-j{\bar \sigma},
\label{potentialrewritten}
}
can be reduced, where we defined $\lambda:=\lambda_1+\lambda_2/3$.
From the gap equation 
\al{
\frac{\df U}{\df {\bar \sigma}}
=\left(m^2\bra{\bar \sigma}\ket-\frac{c}{\sqrt{6}}\bra{\bar \sigma}\ket^2+\lambda\bra{\bar \sigma}\ket^3-j\right)=0,
\label{mpifpirelation}
}
we obtain 
\al{
m^2=m_\pi^2+\frac{c}{2}f_\pi-\frac{3\lambda}{2}f_\pi^2,
}
where we have used $j=m_\pi^2\bra\bar \sigma\ket$ and the fact Eq.\,\eqref{decaycondesate} that the pion decay constant $f_\pi$ is given by the expectation value $f_\pi=\sqrt{2/3}\bra \bar \sigma\ket $.
Then the effective potential Eq.\,\eqref{potentialrewritten} finally is written by $c$ and $\lambda$ as
\al{
U\fn{\bar \sigma}&=
\frac{1}{2}\left(m_\pi^2+\frac{c}{2}f_\pi-\frac{3\lambda}{2}f_\pi^2\right){\bar \sigma}^2
-\frac{c}{3\sqrt{6}}{\bar \sigma}^3+\frac{\lambda}{4}{\bar \sigma}^4 - \sqrt{\frac{3}{2}}m_\pi^2f_\pi {\bar \sigma}.
}

\begin{figure}
\begin{center}
\includegraphics[width=80mm]{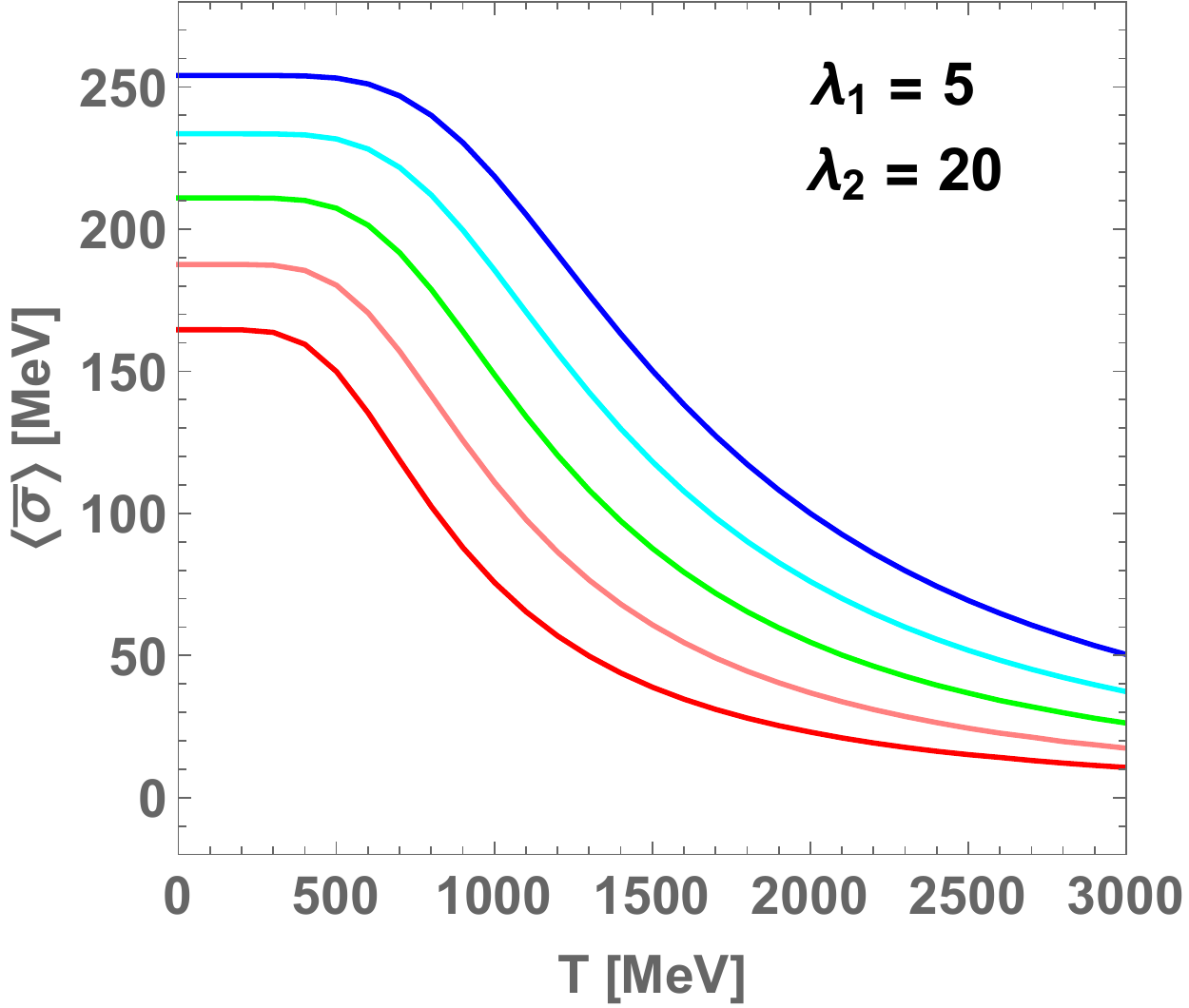}
\includegraphics[width=80mm]{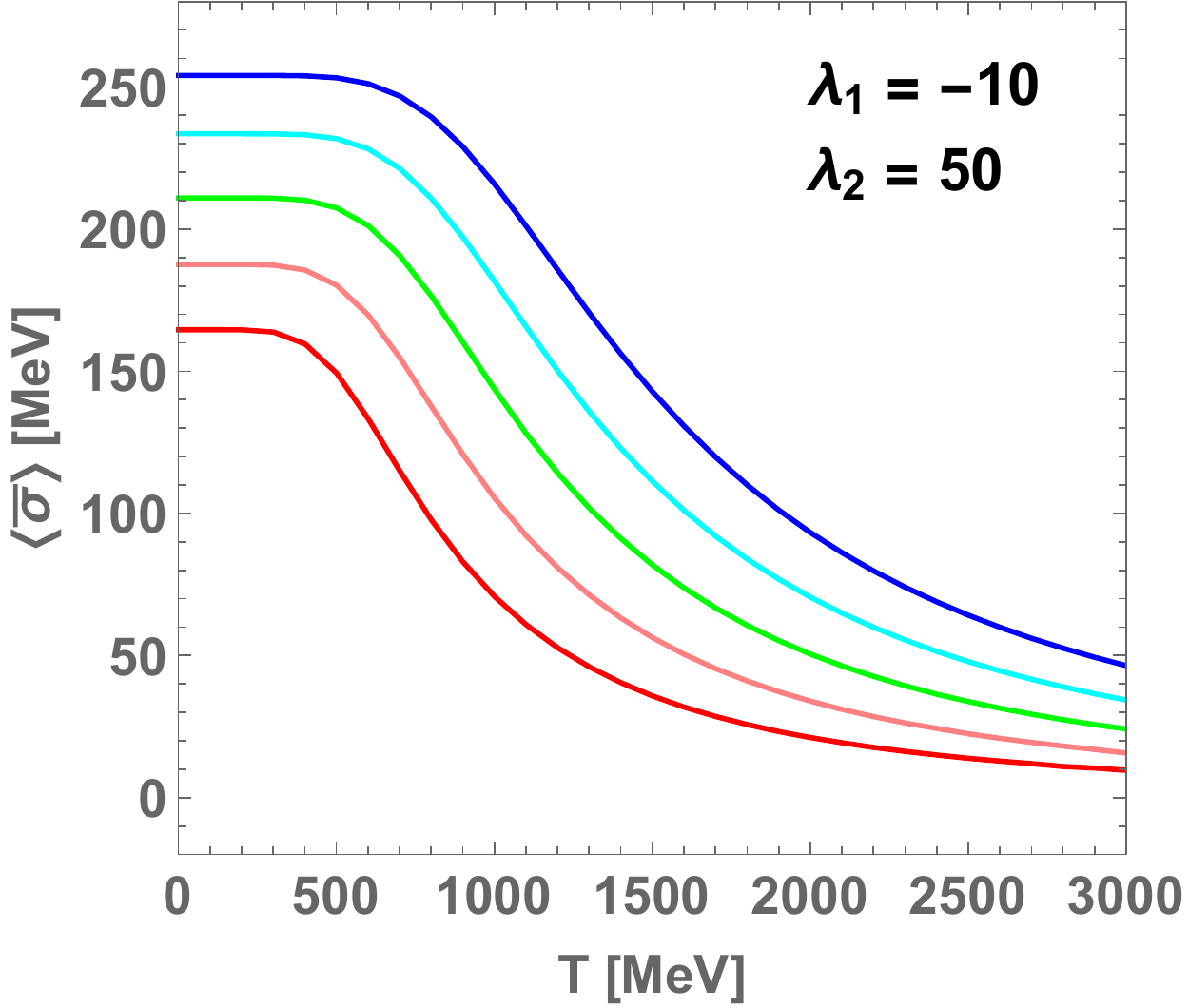}
\end{center}
\begin{center}
\includegraphics[width=80mm]{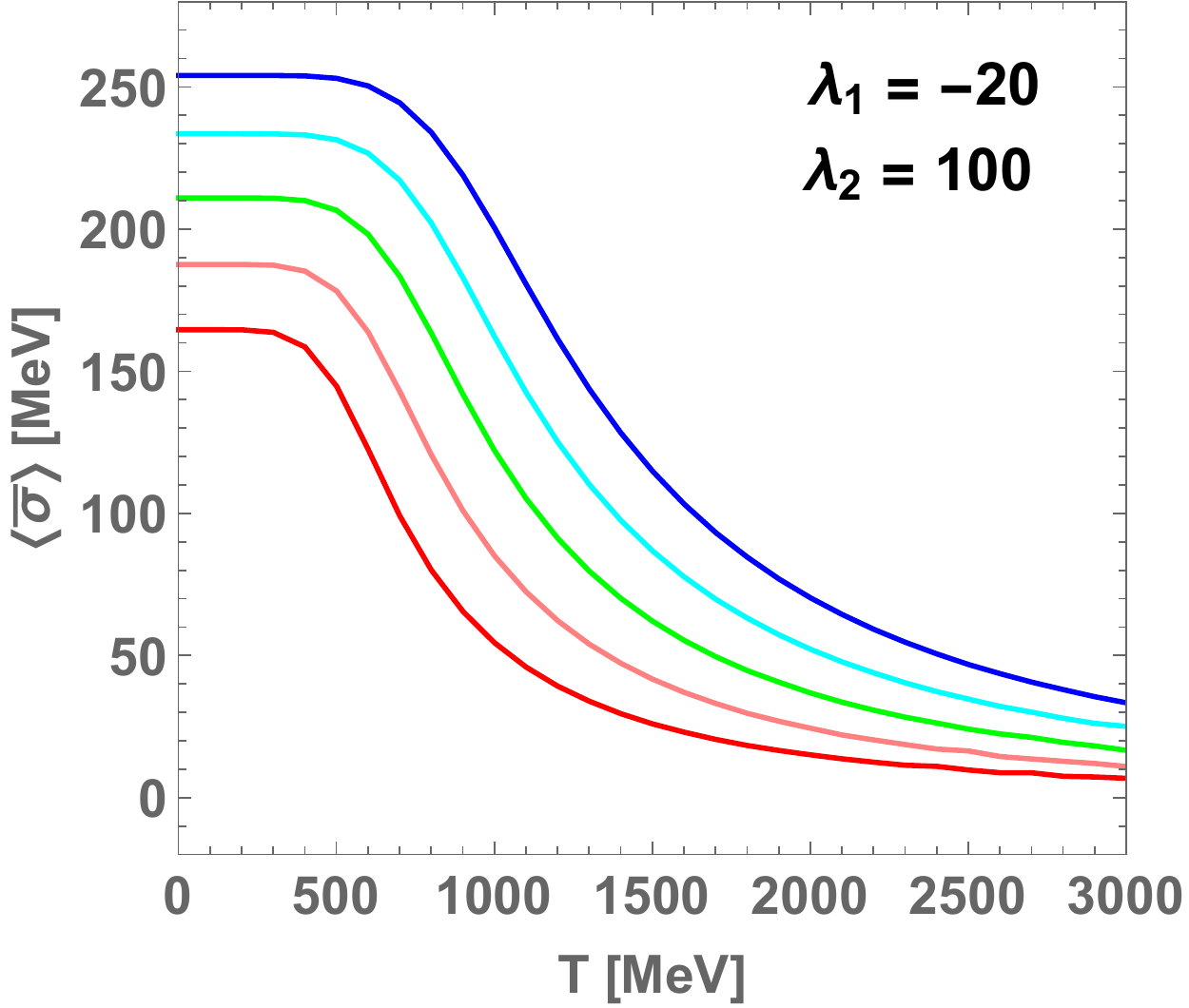}
\includegraphics[width=80mm]{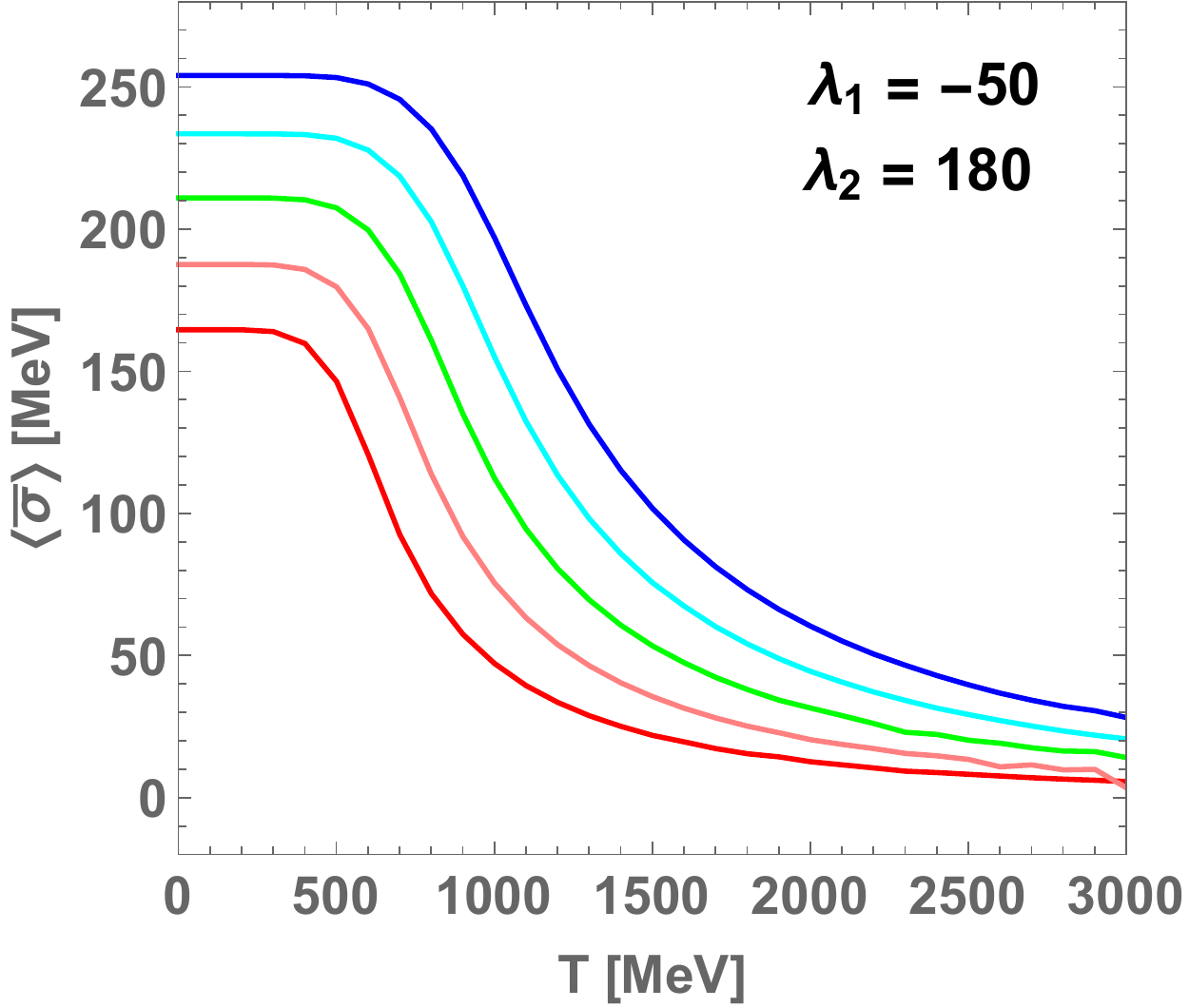}
\end{center}
\vspace{-0.5cm}
\caption{The behavior of the expectation value $\bra \bar \sigma \ket$ for varying temperature.
In each panel, we have taken different $\lambda_1$ and $\lambda_2$ values,
 which are shown therein.
The red, pink, green, cyan, and blue lines correspond to $m_\pi = 1.43$, 1.72, 2.02, 2.31, and 2.60 GeV, respectively.
}
\label{SIMP_crossover}
\end{figure}
Here, let us evaluate the chiral phase transition with vanishing anomaly coupling constant $c=0$.
Input values of $\lambda_1$ and $\lambda_2$ are shown in each panel.
In Fig.~\ref{SIMP_crossover}, we show $\bra \bar \sigma\ket$--$T$ plots with varying $m_\pi$ and $\lambda$ ($\lambda_1$ and $\lambda_2$).
The red, pink, green, cyan, and blue lines correspond to $m_\pi = 1.43$, 1.72, 2.02, 2.31, and 2.60 GeV, respectively.
The values of $f_\pi$ is determined by requirement of the SIMP dark matter,
 i.e. red line of Fig~\ref{allowed region}.
One cannot see the first-order phase transition in any case.
It is reported in Ref.~\cite{Lenaghan:2000ey} that in the absence of the anomaly term the crossover region is narrower than in the presence of the anomaly term.\footnote{
Since as can be seen in Eq.\,\eqref{potentialrewritten} the $\text{U}\fn{1}_A$ anomaly term becomes a cubic coupling in the effective potential for $N_f=3$, it drives the phase transition first-order in the chiral limit.
In the case without the anomaly and quark masses, the first-order phase transition is induced by the fluctuation,  which one can see in the high temperature expansion of Eq.\,\eqref{thermaleffects}.
Note that for the case of $N_f=2$ massless quarks where the $\text{U}\fn{1}_A$ anomaly term is a quadratic coupling, the phase transition becomes first-order in the absence of the anomaly.
}
It it expected that the phase transition remains the crossover even if a finite $c$ is put.\footnote{
In studies of the linear sigma model with $2+1$ flavors of finite quark masses (quark-meson model) by using the mean-field approximation~\cite{Schaefer:2008hk} and the functional renormalization group~\cite{Mitter:2013fxa}, we also see that the $\text{U}\fn{1}_A$ anomaly hardly modify the behavior of the phase transition.
}
Therefore, we conclude that it is difficult to realize the strongly first-order phase transition within the present setup.

\section{Summary}\label{summary}
In this paper,  we have studied the GW background {spectra} produced by the D$\chi$SB in the dark sector with the dark pion as the WIMP or the SIMP dark matter.
In the case where the dark pion is the WIMP dark matter, we have introduced the classically scale invariant model.
We have found that the GW background {spectra} of the sound wave can be observed by the future GW interferometer such as DECIGO.
On the other hand, in the case where the dark pion is the SIMP dark matter, by imposing the constraints for the chiral perturbative limit and {pion-pion scattering} cross section, the chiral phase transition tends to become crossover.
Then the GWs are not produced in this SIMP DM scenario.

The origin of the dark quark mass is essential for making the difference between the WIMP and SIMP {scenario}.
For the former, the dark quark mass is given by the Yukawa interaction.
Since in the thermal history of the phase transitions within the model Eq.\,\eqref{darksectorLag}, the chiral phase transition temperature is higher than the ones of the Higgs and the singlet scalar, the dark quark is massless when the D$\chi$SB takes place.
Therefore, the first-order phase transition can be realized.
In contrast, for the latter, our analysis has treated dark quark masses as finite constant values in the thermal history.\footnote{
The origin of finite dark quark masses could be explained by the non-chiral fixed point in asymptotic safe scenario of quantum gravity~\cite{Eichhorn:2016vvy}; see also the related papers e.g.~\cite{Niedermaier:2006wt,Niedermaier:2006ns,Codello:2008vh,Reuter:2012id,Oda:2015sma,Wetterich:2016uxm,Hamada:2017rvn}.
}
This is the reason why the chiral phase transition becomes crossover.
Note that the classically scale invariant model Eq.\,\eqref{darksectorLag} may be unnatural for the SIMP scenario since the energy scale of the dark sector is much smaller than the EW energy scale.
If we construct a realistic model explaining the dark quark, it may be able to realize the first-order phase transition producing the observable {spectra} of the GW background.

We have found that the GW spectrum from the sound wave has a peak at $\nu\simeq {\mathcal O}\fn{10^{-1}}\, \text{Hz}$ in a scale invariant extension of the WIMP scenario.
In several extended models of the Higgs sector with the phase transitions~\cite{Iso:2009ss,Kubo:2014ova,Kubo:2015cna,Kubo:2015joa,Haba:2015qbz,Ishida:2016ogu,Ishida:2016fbp,Ishida:2017ehu,Haba:2017wwn,Haba:2017quk}, the peak of GW background {spectra} appear at a similar frequency~\cite{Leitao:2012tx,Kakizaki:2015wua,Jinno:2015doa,Jaeckel:2016jlh,Hashino:2016rvx,Jinno:2016knw,Hashino:2016xoj,Kubo:2016kpb,Balazs:2016tbi,Baldes:2017rcu,Chao:2017vrq}.
In order to clarify the origin of the phase transition producing the GWs, it is important to probe the signals from new particles by the {complementary} methods such as the collider experiment and the dark matter direct detection.

Finally we comment on issues which are beyond the scope of this paper.
We need more precise calculations by using, e.g. functional renormalization group and lattice simulation for the dQCD sector.
We should precisely evaluate the effective potential at finite temperature in order to see the cosmological phase transition temperature, the latent heat and the duration time.
Besides, it is important in the WIMP scenario to evaluate the effective couplings $\kappa_s$ and $\kappa_t$ given in Eq.\,\eqref{kappas} and Eq.\,\eqref{kappat}, respectively.
In this work, we have focused on the spontaneous breaking pattern $\text{SU}\fn{3}_R\times \text{SU}\fn{3}_L\to \text{SU}\fn{3}_V$.
Other breaking patterns, such as $\text{SU}\fn{2N_f}\to \text{Sp}\fn{2N_f}$, are also possible.
The {spectra} of GWs should be investigated for these cases.
These issues will be discuss in elsewhere.

\subsection*{Acknowledgement}
We thank the workshop ``Progress in Particle Physics 2016"  at Yukawa Institute for Theoretical Physics, Kyoto University
(YITP-W-16-07) where our collaboration was started.
M.\,Y thanks Jisuke Kubo, Mario Mitter, Jan M. Pawlowski and Fabian Rennecke for valuable discussions.
The work of K.\,T is supported by JSPS Grant-in-Aid for Young Scientists (B) (Grants No.\,16K17697), by the MEXT Grant-in-Aid for Scientific Research on Innovation Areas (Grants No.\,16H00868), and by Kyoto University: Supporting Program for Interaction-based Initiative Team Studies (SPIRITS).
The work of M.\,Y is supported by the DFG Collaborative Research Centre SFB 1225 (ISOQUANT).
The work of Y.\,Y is supported by Research Fellowships of the Japan Society for the Promotion of Science for Young Scientists (Grants No.\,26$\cdot$2428).

\begin{appendix}
\section{{Linear} sigma model}\label{explicit linier sigma model}
For the dQCD sector Eq.\,\eqref{original Lag}, we introduce the $\U\fn{3}_L\times \U\fn{3}_R$ linear sigma model with the axial anomaly term~\cite{Schechter:1971qa,Carruthers:1972ym,Pagels:1974se,Ishida:1999qk,Lenaghan:2000ey,Roder:2003uz}.
In this section, we show its Lagrangian and masses of mesons.
Furthermore, the effective potential at finite temperature is shown.
\subsection{Brief sketch for derivation of meson model from dQCD}
Before showing the Lagrangian of the linear sigma model, we start with briefly sketching out the derivation of the meson model from the dQCD action,
\al{
S=\int \df^4x \left[ -\frac{1}{4}F_{\mu\nu}^aF^a{}^{\mu\nu} + \bar \psi_i (\delta _{ij}i\Slash D - (m_q)_{ij}) \psi_j\right].
}
Now, let us consider integrating out the gauge field.
The effective action would be generated by the dynamics of gauge fields such that
\al{
S_\text{eff}=\int \df^4x   
\left[ Z_\psi \bar \psi_i (\delta _{ij}i\Slash \p - (m_q)_{ij}) \psi_j - 2G\, \tr\left(\chi^\dagger \chi \right)+G_D\left(\det\, \chi + \det\, \chi^\dagger \right) +\cdots \right],
\label{fermionicaction}
}
where ``tr"  and ``det" act on the flavor space, we defined
\al{
\chi_{ij}&=\bar \psi_i (1-\gamma^5)\psi_j =\frac{1}{2}\lambda^a_{ji}\tr\, \psi \lambda^a(1-\gamma^5)\psi,\nn
(\chi^\dagger)_{ij}&=\bar \psi_i (1+\gamma^5)\psi_j =\frac{1}{2}\lambda^a_{ji}\tr\, \psi \lambda^a(1+\gamma^5)\psi,
}
with the generators $\lambda^a$ of $\text{U}\fn{N_f}$ flavor symmetry, and the last term breaks $\U\fn{1}_A$ symmetry due to the anomaly~\cite{Kobayashi:1970ji,tHooft:1976snw,tHooft:1986ooh}.\footnote{
See also \cite{Pawlowski:1996ch} for the treatment of $\text{U}\fn{1}_A$ term.
}
The Lagrangian Eq.\,\eqref{fermionicaction} is invariant under $\SU\fn{N_f}_L\times \SU\fn{N_f}_R\times \U\fn{1}_V$ transformation.\footnote{
Strictly speaking, the $\U\fn{N_f}_A$ symmetry is broken to $Z\fn{N_f}_A$ symmetry by the anomaly~\cite{Pisarski:1983ms}.
However, it is not relevant for the low-energy dynamics.
}
Hereafter, for simplicity we treat only the scalar-type operator $\bar \psi \psi$ with one flavor, and ignore the corrections to the field renormalization $Z_\psi\simeq 1$ and the anomaly term.
That is, we give
\al{
S_\text{eff}\simeq \int \df^4x   
\left[ \bar \psi (i\Slash \p - m_q) \psi - \frac{G}{2}(\bar \psi\psi)^2 \right].
\label{NJLtype}
}
This is the Nambu--Jona-Lasinio model~\cite{Nambu:1961tp} without the pseudo scalar operator $(\bar \psi i\gamma^5 \psi)^2$.

The following Gaussian integral is inserted into the path-integral:
\al{
1={\mathcal N}\int {\mathcal D}\sigma \exp\fn{\frac{i}{2}({\mathcal A}\sigma - {\mathcal B}(\bar \psi\psi) + {\mathcal C})^2},
}
where ${\mathcal N}$, ${\mathcal A}$, ${\mathcal B}$ and ${\mathcal C}$ are constants.
We have the effective action
\al{
S_\text{eff}
&=\int \df^4x   
\left[ \bar \psi (i\Slash \p - m_q) \psi - \frac{G}{2}(\bar \psi\psi)^2 -\frac{1}{2}({\mathcal A}\sigma -{\mathcal B}(\bar \psi\psi) + {\mathcal C})^2  \right]\nn
&=\int \df^4x
\Bigg[ \bar \psi i\Slash \p\psi - \left(m_q-{\mathcal B}{\mathcal C}\right)\bar \psi\psi -\left( \frac{G}{2}-\frac{{\mathcal B}^2}{2}  \right)(\bar \psi\psi)^2 
-{\mathcal A}{\mathcal C}\sigma -\frac{{\mathcal A}^2}{2}\sigma^2 -{\mathcal A}{\mathcal B}\sigma {\bar \psi}\psi
 \Bigg],
 \label{bosonized action}
}
where the constant ${\mathcal C}^2$ is neglected.
Defining 
\al{
{\mathcal B}{\mathcal C}&=m_q,&
{\mathcal B}^2 &=G,&
\label{relation1}
}
the bare quark mass and the four-Fermi interaction terms vanish. 
We also define the new coupling constants as
\al{
y_\sigma&={\mathcal A}{\mathcal B},&
j&={\mathcal A}{\mathcal C},&
m^2&={\mathcal A}^2.&
\label{relation2}
}
Using Eq.\,\eqref{relation1} and Eq.\,\eqref{relation2}, we obtain the relations,
\al{
j&=\frac{m^2m_q}{y_\sigma},&
m^2&=\frac{y_\sigma^2}{G}.
\label{relation3}
}
Note that if we redefine $\sigma \to y_\sigma^{-1} \sigma$ (or equivalently ${\mathcal A}{\mathcal B}=1$), the Yukawa coupling constant disappears.
This means that the Yukawa coupling constant is redundant and then should not determine the low energy physics if we start from the action Eq.\,\eqref{NJLtype}.
This fact actually can be seen in the works using the functional renormalization group with the rebosonization~\cite{Mitter:2014wpa,Braun:2014ata}.

Since in the bosonized action Eq.\,\eqref{bosonized action} the fermionic operators are already bilinear forms, the Gaussian integral can be performed.
Then we obtain
\al{
S_\text{eff}
=\int \df^4x
\Bigg[  
-j\sigma -\frac{m^2}{2}\sigma^2 
-\text{Tr}\log (i{\Slash k}-y_\sigma \sigma)
 \Bigg].
}
The last term corresponds to the loop effects of fermions.
Expanding it into the polynomial of $\sigma$, we obtain the higher order powers of $\sigma$, such as $\sigma^4$.
The kinetic term of $\sigma$ is also generated by the loop effects of fermion via the two-point function:
\al{
\frac{Z_\sigma}{2}= \frac{\df\Gamma^{(2)}\fn{p^2}}{\df p^2}\Bigg|_{p=0},
}
where $Z_\sigma$ is the field renormalization factor of $\sigma$.
As we have sketched, the effective action written in terms of meson is generated from the dQCD action.
At the low scale after the D$\chi$SB, the chiral dynamics could be described by the mesons.
Therefore, we can assume the effective Lagrangian of the dQCD as
\al{
\Lag_\text{eff}=\frac{1}{2}(\p_\mu \sigma)^2 -\frac{m^2}{2}\sigma^2-\frac{\lambda}{4}\sigma^4.
}
Although in principle, the parameters $m^2$ and $\lambda$ are determined by the gauge coupling of  dQCD, it is difficult to precisely determine them due to uncertainties arising from approximations.
Then the effective coupling constants $m^2$ and $\lambda$ are regarded as free parameters in the effective model approach.
The discussion given above can be generalized to the case where the flavor symmetry, the pseudo meson and the anomaly term are involved. 

\subsection{Explicit forms of linear sigma model}
\subsubsection{Lagrangian}
In this paper, we use the following Lagrangian:
\al{
\Lag_\text{LS} &=\tr\left(\p_\mu \Phi \p^\mu \Phi \right)
-m^2\,\tr\left( \Phi^\dagger \Phi \right) -\lambda_1 \left[\tr\left( \Phi^\dagger \Phi\right) \right]^2 
-\lambda_2 \tr\left( \Phi^\dagger \Phi \right) ^2 +c \left( \det\, \Phi +\det\, \Phi^\dagger \right).
\label{effLSm}
}
This Lagrangian is invariant under $\SU\fn{3}_L\times \SU\fn{3}_R\times \U\fn{1}_V$ transformation.
Note that the coupling constant $c$ given in Eq.\,\eqref{effLSm} has to be positive since the coefficient of the instanton-induced six-fermi interaction in the $N_f=3$ case is negative~\cite{tHooft:1976snw,tHooft:1986ooh}.
The field $\Phi$ is defined as
\al{
\Phi =T_a \phi_a =T_a (\sigma_a+i\pi_a),
}
where $\sigma_a$ are the scalar fields, $\pi_a$ are the pseudoscalar fields.
The matrices $T_a=\hat \lambda_a/2$ ($a=0,\cdots, 8$) are the generator of $\U\fn{3}$ with $\lambda_0=\sqrt{\frac{2}{3}}I$ and the Gell-mann matrices $\hat \lambda_a$ for $a=1,\cdots, 8$.
They satisfy the identities 
\al{
[T_a,T_b]&=if_{abc}T_c,& 
\{ T_a, T_b\} &= d_{abc}T_c,&
\text{Tr}(T_aT_b)&=\frac{\delta_{ab}}{2},&
\label{matrixformula}
}
where $f_{abc}$ and $d_{abc}$ are the antisymmetric and symmetric structure constants of $\SU\fn{3}$ for $a,\, b,\, c=1,\cdots,8$ and 
\al{
f_{ab0}&=0,&d_{ab0}&=\sqrt{\frac{2}{3}}\delta_{ab}.&
}
$\Phi=(\Phi)_{ij}$ is the scalar field transformed as
\al{
\Phi \to U_L \Phi U_R^\dagger,
}
where $U_{L(R)} =\exp \fn{i\theta^a_{L(R)}T_a}$.
Note that 
\al{
\tr\,[T_a]=\delta_{a0}\sqrt{\frac{2}{3}}\tr\,\fn{\frac{I}{2}}=\sqrt{\frac{3}{2}}\delta_{a0}.
}

The Lagrangian in terms of the fields $\sigma^a$ and $\pi^a$ becomes
\al{
\Lag_\text{LS}
&=\frac{1}{2}\left[\p_\mu \sigma_a\p^\mu \sigma_a+\p_\mu \pi_a\p^\mu \pi_a-\sigma_a(m^2\delta_{ab})\sigma_b-\pi_a(m^2\delta_{ab})\pi_b\right]\nn
&\quad +{\mathcal G}_{abc}\sigma_a\sigma_b\sigma_c
 -3{\mathcal G}_{abc} \pi_a\pi_b\sigma_c
 -2{\mathcal H}_{abcd}\sigma_a\sigma_b\pi_c\pi_d
  -\frac{1}{3}{\mathcal F}_{abcd}(\sigma_a\sigma_b\sigma_c\sigma_d+\pi_a\pi_b\pi_c\pi_d),
}
where we have defined 
\al{
{\mathcal G}_{abc}&:=\frac{c}{6} 
\left[d_{abc}
-\frac{3}{2}(\delta_{a0}d_{0bc}+\delta_{b0}d_{a0c}+\delta_{c0}d_{ab0})
+\frac{9}{2}d_{000}\delta_{a0}\delta_{b0}\delta_{c0}
\right],
\label{Gabc}
\\
{\mathcal F}_{abcd}&:= \frac{\lambda_1}{4}(\delta_{ab}\delta_{cd}+\delta_{ad}\delta_{bc}+\delta_{ac}\delta_{bd})
+\frac{\lambda_2}{8}(d_{abn}d_{ncd}+d_{adn}d_{nbc}+d_{acn}d_{nbd}),
\label{Fabcd}
\\
{\mathcal H}_{abcd}&:=\frac{\lambda_1}{4}\delta_{ab}\delta_{cd}
+\frac{\lambda_2}{8}(d_{abn}d_{ncd}+f_{acn}f_{nbd}+f_{bcn}f_{nad}).
\label{Habcd}
}

After the spontaneous symmetry breaking, the field $\Phi$ obtains a vacuum expectation value as
\al{
\bra \Phi \ket = T_a\bar \sigma_a,
\Rightarrow
\sigma_a \to \sigma_a+\bar \sigma_a.
}
In this case the Lagrangian becomes
\al{
\Lag_\text{LS}&=
\frac{1}{2}\bigg[\p_\mu \sigma_a\p^\mu \sigma_a
+\p_\mu \pi_a\p^\mu \pi_a\nn
&\quad-\sigma_a(m^2\delta_{ab}
-6{\mathcal G}_{abc}\bar \sigma_c
+4{\mathcal F}_{abcd}\bar \sigma_c\bar \sigma_d
)\sigma_b
-\pi_a(m^2\delta_{ab}
+6{\mathcal G}_{abc}\bar \sigma_c
+4{\mathcal H}_{abcd}\bar \sigma_c\bar \sigma_d)\pi_b\bigg]\nn
&\qquad +\bigg({\mathcal G}_{abc}-\frac{4}{3}{\mathcal F}_{abcd}\bar \sigma_d\bigg) \sigma_a\sigma_b\sigma_c
-3 \bigg({\mathcal G}_{abc}+\frac{4}{3}{\mathcal H}_{abcd}\bar \sigma_d\bigg)\pi_a\pi_b\sigma_c\nn
&\quad -2{\mathcal H}_{abcd}\sigma_a\sigma_b\pi_c\pi_d-\frac{1}{3}{\mathcal F}_{abcd}(\sigma_a\sigma_b\sigma_c\sigma_d+\pi_a\pi_b\pi_c\pi_d)\nn
&\qquad + \frac{1}{2}\bar \sigma_a(m^2\delta_{ab})\bar \sigma_b
-{\mathcal G}_{abc}\bar \sigma_a\bar \sigma_b\bar \sigma_c
+\frac{1}{3}{\mathcal F}_{abcd}\bar \sigma_a\bar \sigma_b\bar \sigma_c\bar \sigma_d.
\label{lagrangianbroken}
}

The explicit breaking term corresponding to the quark masses is introduced as
\al{
\Lag_\text{SB} = \text{Tr}[H(\Phi +\Phi^\dagger)]=j_a\sigma_a,
}
with $H=T_a j_a$.
This breaking term is given as
\al{
\Lag_\text{SB} &= \text{Tr}\left[ \left(j_0T_0+j_3 T_3+j_8T_8 \right)\left(\Phi+\Phi^\dagger\right)\right]\nn
&=j_0\sigma_0+j_3\sigma_3+j_8\sigma_8,
}
where $j_i$ are proportional to the current dark quark masses $m_{u,d,s}$ :
\al{
j_0 &\propto \frac{2}{3}\tr[(T_0)(m_u, m_d, m_s)^T]=\frac{m_u+m_d+m_s}{3},&\\
j_3 &\propto \tr[(T_3)(m_u, m_d, m_s)^T]=\frac{m_u-m_d}{2},&\\
j_8 &\propto \frac{1}{\sqrt{3}} \tr[(T_8)(m_u, m_d, m_s)^T]=\frac{m_u+m_d-2m_s}{6},&
}
In this work, for simplicity we assume the dark quark masses to be degenerate, $m_u=m_d=m_s$.
In this case, $j_0\propto m_u$ and $j_3 =j_8=0$, and $\bra \sigma \ket=T_0\bar \sigma_0$.
The effective potential at the tree level becomes
\al{
U\fn{\bar \sigma_0}
&=\frac{m^2}{2}\bar \sigma_0^2
-\frac{c}{3\sqrt{6}}\bar \sigma_0^3
+\frac{1}{4}\left( \lambda_1 + \frac{\lambda_2}{3} \right){\bar \sigma}_0^4
-j_0\bar \sigma_0.
\label{tree level potential}
}

To summarize, the Lagrangian is rewritten as
\al{
\Lag&= 
\Lag_\text{LS}+\Lag_\text{SB}\nn
&=\frac{1}{2}[
\p_\mu \sigma_a \p^\mu \sigma_a
+\p_\mu \pi \p^\mu \pi_a
-\sigma_a(m^2_S)_{ab}\sigma_b 
- \pi_a(m^2_P)_{ab}\pi_b]
+\bigg({\mathcal G}_{abc}-\frac{4}{3}{\mathcal F}_{abcd}\bar \sigma_d\bigg) \sigma_a\sigma_b\sigma_c\nn
&\quad -3 \bigg({\mathcal G}_{abc}+\frac{4}{3}{\mathcal F}_{abcd}\bar \sigma_d\bigg)\pi_a\pi_b\sigma_c
-2{\mathcal H}_{abcd}\sigma_a\sigma_b\pi_c\pi_d-\frac{1}{3}{\mathcal F}_{abcd}(\sigma_a\sigma_b\sigma_c\sigma_d+\pi_a\pi_b\pi_c\pi_d)
-U\fn{\bar \sigma},
}
where the mass terms are
\al{
\sigma_a(m^2_S)_{ab}\sigma_b + \pi_a(m^2_P)_{ab}\pi_b&:=\sigma_a[m^2 \delta_{ab} -6{\mathcal G}_{abc}\bar \sigma_c+4{\mathcal F}_{abcd}\bar \sigma_c \bar \sigma_d]\sigma_b,\nn
&\quad+\pi_a[m^2 \delta_{ab} + 6{\mathcal G}_{abc}\bar \sigma_c+4{\mathcal H}_{abcd}\bar \sigma_c \bar \sigma_d]\pi_b,
}
and the potential at the tree level is
\al{
U\fn{\bar \sigma}=\frac{m^2}{2}\bar \sigma_a^2-{\mathcal G}_{abc}\bar \sigma_a\bar \sigma_b\bar \sigma_c +\frac{1}{3}{\mathcal F}_{abcd} \bar \sigma_a\bar \sigma_b\bar \sigma_c\bar \sigma_d -j_a\bar \sigma_a.
\label{generaltreepotential}
}
The expectation value $\bar \sigma$ at the tree level is determined by
\al{
\frac{\p U\fn{\bar \sigma}}{\p \bar \sigma_a}
=m^2 \bar \sigma_a -3{\mathcal G}_{abc}\bar\sigma_b\bar\sigma_c
+\frac{4}{3}{\mathcal F}_{abcd}\bar\sigma_b\bar\sigma_c\bar\sigma_d-j_a=0.
}

\subsubsection{Tree potential}
For $a,b,c=0,3,8$,
\al{
{\mathcal G}_{abc}&:=\frac{c}{6} 
\left[d_{abc}
-\frac{3}{2}(\delta_{a0}d_{0bc}+\delta_{b0}d_{a0c}+\delta_{c0}d_{ab0})
+\frac{9}{2}d_{000}\delta_{a0}\delta_{b0}\delta_{c0}
\right],\\
{\mathcal F}_{abcd}&:= \frac{\lambda_1}{4}(\delta_{ab}\delta_{cd}+\delta_{ad}\delta_{bc}+\delta_{ac}\delta_{bd})
+\frac{\lambda_2}{8}(d_{abn}d_{ncd}+d_{adn}d_{nbc}+d_{acn}d_{nbd}),
}
we have
\al{
U\fn{\bar \sigma}&=\frac{m^2}{2}(\bar \sigma_0^2+\bar \sigma_3^2+\bar \sigma_8^2)
-\frac{\lambda_2}{3\sqrt{2}}\bar \sigma_0\bar \sigma_8^3
-\frac{c}{3\sqrt{6}}\left(\bar \sigma_0^3+\frac{\bar \sigma_3^3}{2}\right)\nn
&\quad +\left( \frac{\lambda_1}{4}+\frac{\lambda_2}{12}\right) \bar \sigma_0^4
+\left( \frac{\lambda_1}{4}+\frac{\lambda_2}{8}\right) \bar \sigma_3^4
+\left( \frac{\lambda_1}{4}+\frac{\lambda_2}{8}\right) \bar \sigma_8^4\nn
&\quad -\frac{c}{2\sqrt{6}}(\bar \sigma_3^2+\bar \sigma_8^2)\bar \sigma_0
+\frac{1}{2}(\lambda_1+\lambda_2)(\bar \sigma_3^2+\bar \sigma_8^2)\bar \sigma_0^2
-\frac{c}{2\sqrt{3}}\bar \sigma_3^2\bar \sigma_8\nn
&\quad +\frac{1}{2}\left( \lambda_1+\frac{1}{2}\lambda_2\right)\bar \sigma_3^2\bar \sigma_8^2
+\frac{\lambda_2}{\sqrt{2}}\bar \sigma_0\bar \sigma_3^2\bar \sigma_8
-j_0\bar \sigma_0-j_3\bar \sigma_3-j_8\bar \sigma_8.
}

Note that the gap equations are given by
\al{
\frac{\p U}{\p \bar \sigma_0}=0\iff
j_0&=\left[m^2 -\frac{c}{\sqrt{6}}\bar \sigma_0+\left(\lambda_1+\frac{\lambda_2}{3}\right)\sigma_0^2\right]\bar \sigma_0\nn
&\quad +\left[
\frac{c}{2\sqrt{6}}+(\lambda_1+\lambda_2)\bar \sigma_0 -\frac{\lambda_2}{3\sqrt{2}}\bar \sigma_8
\right]\bar \sigma_8^2\nn
&\qquad +\left( \frac{c}{2\sqrt{6}} +\frac{\lambda_2}{\sqrt{2}}\bar \sigma_8\right)\bar \sigma_3^2
+(\lambda_1+\lambda_2)\bar \sigma_0\bar \sigma_3^2,
\label{gapequ0}
\\
\frac{\p U}{\p \bar \sigma_8}=0\iff
j_8&=\bigg[ 
m^2+\frac{c}{\sqrt{6}}\bar \sigma_0+\frac{c}{2\sqrt{3}}\bar \sigma_8
+(\lambda_1+\lambda_2)\bar \sigma_0^2-\frac{\lambda_2}{\sqrt{2}}\bar \sigma_0\bar \sigma_8+\left(\lambda_1+\frac{\lambda_2}{2}\right)\bar \sigma_8^2
\bar \sigma_8\nn
&\quad -\frac{c}{2\sqrt{3}}
+\frac{\lambda_2}{\sqrt{2}}\bar \sigma_0 
+\left( \lambda_1 +\frac{\lambda_2}{2} \right)\bar \sigma_3^2\bigg]\bar \sigma_8,\\
\frac{\p U}{\p \bar \sigma_3}=0\iff
j_3&=\bigg[ 
m^2 +\frac{c}{\sqrt{6}}\bar \sigma_0 -\frac{c}{\sqrt{3}}\bar \sigma_8
+(\lambda_1+\lambda_2)\bar \sigma_0^2
+\sqrt{2}\lambda_2\bar \sigma_0\bar \sigma_8
+\left(\lambda_1+\frac{\lambda_2}{2}\right)\bar \sigma_8^2\nn
&\quad +\left(\lambda_1+\frac{\lambda_2}{2}\right)\bar \sigma_3^2
\bigg]\bar \sigma_3,
}
\subsubsection{Mass {spectra}}
The mass matrix for sigma mesons is given by
\al{
(m^2_S)_{ab}=m^2 \delta_{ab} -6{\mathcal G}_{abc}\bar \sigma_c+4{\mathcal F}_{abcd}\bar \sigma_c \bar \sigma_d.
\label{Smass}
}
That for pseudo-scalar mesons is
\al{
(m^2_P)_{ab}=
m^2 \delta_{ab} + 6{\mathcal G}_{abc}\bar \sigma_c+4{\mathcal H}_{abcd}\bar \sigma_c \bar \sigma_d.
\label{Pmass}
}
We list the explicit forms for $a=1,3,8$:\\
\al{
(m^2_S)_{00}&=m^2
-\sqrt{\frac{2}{3}}c\, \bar \sigma_0 
+ (3\lambda_1+\lambda_2)\bar \sigma_0^2+(\lambda_1+\lambda_2)\bar \sigma_3^2+(\lambda_1+\lambda_2)\bar \sigma_8^2,\\
(m^2_S)_{11}&=(m^2_S)_{22}\nn
&=m^2
+\frac{c}{\sqrt{6}}\bar \sigma_0
-\frac{c}{\sqrt{3}}\bar \sigma_8
+(\lambda_1+\lambda_2)\bar \sigma_0^2 
+ \left(\lambda_1+\frac{\lambda_2}{2}\right) \bar \sigma_3^2
+\sqrt{2}\lambda_2 \bar \sigma_0\bar \sigma_8 
+\left(\lambda_1+\frac{\lambda_2}{2}\right) \bar \sigma_8^2,\\
(m^2_S)_{33}&=m^2
+\frac{c}{\sqrt{6}}\bar \sigma_0
-\frac{c}{\sqrt{3}}\bar \sigma_8
+(\lambda_1+\lambda_2)\bar \sigma_0^2 
+ 3\left(\lambda_1+\frac{\lambda_2}{2}\right) \bar \sigma_3^2
+\sqrt{2}\lambda_2 \bar \sigma_0\bar \sigma_8 
+\left(\lambda_1+\frac{\lambda_2}{2}\right) \bar \sigma_8^2,\\
(m^2_S)_{44}&=(m^2_S)_{55}\nn
&=m^2
+\frac{c}{\sqrt{6}}\bar \sigma_0
-\frac{c}{2}\bar \sigma_3
+\frac{c}{2\sqrt{3}}\bar \sigma_8
+(\lambda_1+\lambda_2)\bar \sigma_0^2 
+ \left(\lambda_1+\frac{\lambda_2}{2}\right) \bar \sigma_3^2
+\sqrt{\frac{3}{2}}\lambda_2\bar \sigma_0\bar \sigma_3\nn
&\quad -\frac{\lambda_2}{\sqrt{2}} \bar \sigma_0\bar \sigma_8 
+\left(\lambda_1+\frac{\lambda_2}{2}\right) \bar \sigma_8^2,\\
(m^2_S)_{66}&=(m^2_S)_{77}\nn
&=m^2
+\frac{c}{\sqrt{6}}\bar \sigma_0
+\frac{c}{2}\bar \sigma_3
+\frac{c}{2\sqrt{3}}\bar \sigma_8
+(\lambda_1+\lambda_2)\bar \sigma_0^2 
+ \left(\lambda_1+\frac{\lambda_2}{2}\right) \bar \sigma_3^2
-\sqrt{\frac{3}{2}}\lambda_2\bar \sigma_0\bar \sigma_3\nn
&\quad -\frac{\lambda_2}{\sqrt{2}} \bar \sigma_0\bar \sigma_8 
+\left(\lambda_1+\frac{\lambda_2}{2}\right) \bar \sigma_8^2,\\
(m^2_S)_{88}&=m^2
+\frac{c}{\sqrt{6}}\bar \sigma_0
+\frac{c}{\sqrt{3}}\bar \sigma_8
+(\lambda_1+\lambda_2)\bar \sigma_0^2 
+ \left(\lambda_1+\frac{\lambda_2}{2}\right) \bar \sigma_3^2
-\sqrt{2}\lambda_2 \bar \sigma_0\bar \sigma_8 
+3\left(\lambda_1+\frac{\lambda_2}{2}\right) \bar \sigma_8^2,\\
(m^2_S)_{30}&=(m^2_S)_{03}
=\left[\frac{c}{\sqrt{6}} +2(\lambda_1+\lambda_2)\bar \sigma_0+\sqrt{2}\lambda_2\bar \sigma_8 \right]\bar \sigma_3,\\
(m^2_S)_{80}&=(m^2_S)_{08}
=\left[\frac{c}{\sqrt{6}} 
+2(\lambda_1+\lambda_2)\bar \sigma_0
-\frac{\lambda_2}{\sqrt{2}}\bar \sigma_8 \right]\bar \sigma_8
+\frac{\lambda_2}{\sqrt{2}}\bar \sigma_3^2,\\
(m^2_S)_{83}&=(m^2_S)_{38}
=\left[-\frac{c}{\sqrt{3}} +\sqrt{2}\lambda_2\bar \sigma_0 
+2\left( \lambda_1+\frac{\lambda_2}{2} \right) \bar \sigma_8
 \right]\bar \sigma_3,\\
(m^2_P)_{00}&=
m^2 
+\sqrt{\frac{2}{3}}c \bar \sigma_0
+\left( \lambda_1 +\frac{\lambda_2}{3}\right)\bar \sigma_0^2
+\left( \lambda_1 +\frac{\lambda_2}{3}\right)\bar \sigma_3^2
+\left( \lambda_1 +\frac{\lambda_2}{3}\right)\bar \sigma_8^2,\\
(m^2_P)_{11}&=(m^2_P)_{22}\nn
&=m^2 
-\frac{c}{\sqrt{6}}\bar \sigma_0
+\frac{c}{\sqrt{3}}\bar \sigma_8\nn
&\quad +\left( \lambda_1 +\frac{\lambda_2}{3}\right)\bar \sigma_0^2
+\left( \lambda_1 +\frac{3\lambda_2}{2}\right)\bar \sigma_3^2
+\left( \lambda_1 +\frac{\lambda_2}{6}\right)\bar \sigma_8^2
+\frac{\sqrt{2}\lambda_2}{3}\bar \sigma_0\bar \sigma_8,
\label{pionmass}
\\
(m^2_P)_{33}&=
m^2
-\frac{c}{\sqrt{6}}\bar \sigma_0
+\frac{c}{\sqrt{3}}\bar \sigma_8\nn
&\quad +\left( \lambda_1 +\frac{\lambda_2}{3}\right)\bar \sigma_0^2
+\left( \lambda_1 +\frac{\lambda_2}{2}\right)\bar \sigma_3^2
+\left( \lambda_1 +\frac{\lambda_2}{6}\right)\bar \sigma_8^2
+\frac{\sqrt{2}\lambda_2}{3}\bar \sigma_0\bar \sigma_8,\\
(m^2_P)_{44}&=(m^2_P)_{55}\nn
&=m^2
-\frac{c}{\sqrt{6}}\bar \sigma_0
+\frac{c}{2}\bar \sigma_3
-\frac{c}{2\sqrt{3}}\bar \sigma_8
+\left( \lambda_1 +\frac{\lambda_2}{3}\right)\bar \sigma_0^2
+\frac{\lambda_2}{\sqrt{6}}\bar \sigma_0\bar \sigma_3\nn
&\quad +\left( \lambda_1 +\frac{\lambda_2}{2}\right)\bar \sigma_3^2
-\frac{\lambda_2}{3\sqrt{2}}\bar \sigma_0\bar \sigma_8
+\frac{2\lambda_2}{\sqrt{3}}\bar \sigma_3\bar \sigma_8
+\left( \lambda_1 + \frac{7\lambda_2}{6}\right)\bar \sigma_8^2,\\
(m^2_P)_{66}&=(m^2_P)_{77}\nn
&=m^2
-\frac{c}{\sqrt{6}}\bar \sigma_0
-\frac{c}{2}\bar \sigma_3
-\frac{c}{2\sqrt{3}}\bar \sigma_8
+\left( \lambda_1 +\frac{\lambda_2}{3}\right)\bar \sigma_0^2
-\frac{\lambda_2}{\sqrt{6}}\bar \sigma_0\bar \sigma_3\nn
&\quad+ \left( \lambda_1 +\frac{\lambda_2}{2}\right)\bar \sigma_3^2
-\frac{\lambda_2}{3\sqrt{2}}\bar \sigma_0\bar \sigma_8
-\frac{2\lambda_2}{\sqrt{3}}\bar \sigma_3\bar \sigma_8
+\left( \lambda_1 + \frac{7\lambda_2}{6}\right)\bar \sigma_8^2,\\
(m^2_P)_{88}&=m^2
-\frac{c}{\sqrt{6}}\bar \sigma_0
-\frac{c}{\sqrt{3}}\bar \sigma_8
+\left( \lambda_1 +\frac{\lambda_2}{3}\right)\bar \sigma_0^2
+ \left( \lambda_1 +\frac{\lambda_2}{6}\right)\bar \sigma_3^2
-\frac{\sqrt{2}\lambda_2}{3}\bar \sigma_0\bar \sigma_8
+\left( \lambda_1 + \frac{\lambda_2}{2}\right)\bar \sigma_8^2,\\
(m^2_P)_{30}&=(m^2_P)_{03}
=\left[-\frac{c}{\sqrt{6}}
+\frac{2\lambda_2}{3}\bar \sigma_0 
+\frac{\sqrt{2}\lambda_2}{3}\bar \sigma_8
\right]\bar \sigma_3,\\
(m^2_P)_{80}&=(m^2_P)_{08}
=-\frac{c}{\sqrt{6}}\bar \sigma_8
+\frac{\lambda_2}{3\sqrt{2}}\bar \sigma_3^2
+\frac{2\lambda_2}{3}\bar \sigma_0\bar \sigma_8
-\frac{\lambda_2}{3\sqrt{2}}\bar \sigma_8^2,\\
(m^2_P)_{83}&=(m^2_P)_{38}
=\left[
\frac{c}{\sqrt{3}}
+\frac{\sqrt{2}\lambda_2}{3}\bar \sigma_0
+\frac{\lambda_2}{3}\bar \sigma_8
\right]\bar \sigma_3.
}

Note that in case of $\bra\sigma\ket=T_0{\bar \sigma}_0$, the gap equation Eq.\,\eqref{gapequ0} is given by 
\al{
j&=\left[m^2 -\frac{c}{\sqrt{6}}\bar \sigma_0+\left(\lambda_1+\frac{\lambda_2}{3}\right)\bar \sigma_0^2\right]\bar \sigma_0.
}
Since the mass of the pseudo scalar at the tree level is
\al{
(m_P^2)_{ab}=
\left[
m^2 
-\frac{c}{\sqrt{6}}\bar \sigma
+\left( \lambda_1 +\frac{\lambda_2}{3}\right)\bar \sigma^2
\right]\delta_{ab}
+\sqrt{\frac{3}{2}}c\, \sigma \delta_{a0}\delta_{b0},
}
it can be rewritten as 
\al{
(m_P^2)_{ab}=\frac{j}{\bar \sigma_0}\delta_{ab}+\sqrt{\frac{3}{2}}c\, \bar \sigma_0 \delta_{a0}\delta_{b0}.
\label{etapionmass}
}
Since $(m_P^2)_{00}$ is heavier than $(m_P^2)_{ab}$ for $a,b=1,...,8$ due to the positive $c$ that corresponds to the $U\fn{1}_A$ anomaly, the pseudo scalar with $(m_P^2)_{00}$ is actually identified with the $\eta'$ meson.
The dark matter has to be stable, that is, lightest particle.
Therefore, $(m_P^2)_{ab}=:m_\pi^2$ ($a,b=1,...,8$) is the dark pion mass which is proportional to $j$;
\al{
m_\pi^2=\frac{j}{\bar \sigma}.
\label{pionmassrelation}
}

\subsubsection{Decay constant}\label{decayconstantderivation}
Consider the partially conserved axial current relation
\al{
\bra 0| J^\mu_a |\pi^a \ket = ip^\mu f_a.
}
Here $J^\mu_a$ is the Neother current for the axial-vector transformation $\Phi\to \Phi +\theta_A^a\{T^a,T^b\}\phi^b$,
\al{
J^\mu_a&=\frac{\delta \Lag_\text{LS}}{\delta(\p_\mu \phi_b)}(id_{abc}\phi_c) +\text{h.c.}\nn
&=d_{abc}(\sigma_b \p^\mu \pi_c -\pi_b \p^\mu \sigma_c).
}
Assuming that the scalar field $\sigma_a$ has the expectation value; $\sigma_a\to \bar \sigma_a +\sigma_a$, we obtain
\al{
f_a=\sum_{b}d_{aab}\bar \sigma_b.
}
In the case $\bar \sigma_3=\bar \sigma_8=0$, it becomes
\al{
f_\pi=f_1=\sqrt{\frac{2}{3}}\bar \sigma_0.
\label{decaycondesate}
}

\subsection{Effective potential at finite temperature}\label{EFPlinear}
To evaluate the chiral phase transition, we consider the linear sigma model at finite temperature.
It is known that due to the infrared divergence, the perturbation theory for this model breaks down~\cite{Dolan:1973qd}.
Therefore, the so-called ring diagram has to be summed up.
Several methods such as the large-$N$ approximation~\cite{Fejos:2012pm} and the functional renormalization group~\cite{Mitter:2013fxa,Fejos:2014qga,Fejos:2016hbp} have been applied.
In this work, we use the Cornwall-Jackiw-Tomboulis (CJT) formalism~\cite{Cornwall:1974vz}.
We here follow Refs.~\cite{Lenaghan:2000ey,Roder:2003uz} where the CJT formalism for the $\text{U}\fn{N_f}_L\times \text{U}\fn{N_f}_R$ linear sigma model was employed to investigate the chiral phase transition at finite temperature in QCD.

The effective potential in the CJT formalism reads
\al{
V_\text{eff}\fn{\bar \sigma,D_S,D_P ;T}
&=U\fn{\bar \sigma} 
+\frac{1}{2}\int _k \left( \left[\log\fn{D_S^{-1}\fn{k}} \right]_{aa} +\left[\log\fn{D_P^{-1}\fn{k}} \right]_{aa}  \right)\nn
&\quad
+\frac{1}{2}\int _k \left( S_{ab}^{-1}\fn{k;\bar \sigma} [D_S\fn{k}]_{ba} +P_{ab}\fn{k;\bar \sigma} [D_P^{-1}\fn{k}]_{ba} -2\delta_{ab}\delta_{ba} \right)\nn
&\qquad
+V_2\fn{\bar \sigma,D_S,D_P}.
\label{fTeffectivepotential}
}
Here we used shorthand notation,
\al{
\int_k f\fn{k}=\int \frac{\df ^4k}{(2\pi)^4}f\fn{k}=T\sum_{n}\int \frac{\df^3k}{(2\pi)^3}f\fn{\omega_n,{\vec k}},
}
$U\fn{\bar \sigma}$ is the tree-level potential given in Eq.\,\eqref{generaltreepotential} and 
\al{
V_2\fn{\bar \sigma,D_S,D_P;T}
&={\mathcal F}_{abcd}\left[ \int_k [D_S\fn{k}]_{ab} \int_p [D_S\fn{p} ]_{cd} 
+ \int_k [D_P\fn{k}]_{ab} \int_p [D_P\fn{p} ]_{cd}
 \right]\nn
&\quad
+2{\mathcal H}_{abcd}\int_k [D_S\fn{k}]_{ab} \int_p [D_P\fn{p} ]_{cd}
}
corresponds to the sum of two-particle irreducible (2PI) diagrams (double-bubble diagrams) where ${\mathcal G}_{abc}$, ${\mathcal F}_{abcd}$ and ${\mathcal H}_{abcd}$ are defined in Eq.\,\eqref{Gabc}, Eq.\,\eqref{Fabcd} and Eq.\,\eqref{Habcd}, respectively.

The tree-level propagators $S_{ab}^{-1}\fn{k;\bar \sigma}$ and $P_{ab}^{-1}\fn{k;\bar \sigma}$ are
\al{
S_{ab}^{-1}\fn{k;\bar \sigma}&=-k^2\delta_{ab} + (m_S^2)_{ab}
\label{treeSpro},\\
P_{ab}^{-1}\fn{k;\bar \sigma}&=-k^2\delta_{ab} + (m_P^2)_{ab},
\label{treePpro}
}
where the tree-level squared masses $(m_S^2)_{ab}$ and $(m_P^2)_{ab}$ are given in Eq.\,\eqref{Smass} and Eq.\,\eqref{Pmass}, respectively.
The full propagators are determined by the stationary conditions,
\al{
\frac{\delta V_\text{eff}\fn{\bar \sigma,D_S,D_P;T}}{\delta [D_S\fn{k}]_{ab}}=0\iff &
[D_S^{-1}\fn{k}]_{ab}= S_{ab}^{-1}\fn{k;\bar \sigma} +\Sigma_{ab}\fn{k},
\label{fullSpro}
\\
\frac{\delta V_\text{eff}\fn{\bar \sigma,D_S,D_P;T}}{\delta [D_P\fn{k}]_{ab}}=0\iff &
[D_P^{-1}\fn{k}]_{ab}= P_{ab}^{-1}\fn{k;\bar \sigma} +\Pi_{ab}\fn{k},
\label{fullPpro}
}
where we defined the self-energies of the scalar and pseudoscalar particles,
\al{
\Sigma_{ab}\fn{k}&=2\frac{\delta V_2\fn{\bar \sigma,D_S,D_P;T}}{\delta [D_S\fn{k}]_{ab}}
=4{\mathcal F}_{abcd}\int_k [D_S\fn{k}]_{cd} +4{\mathcal H}_{abcd}\int_k [D_P\fn{k}]_{cd} 
\label{selfSenergy}
,\\
\Pi_{ab}\fn{k}&=2\frac{\delta V_2\fn{\bar \sigma,D_S,D_P;T}}{\delta [D_P\fn{k}]_{ab}}
=4{\mathcal H}_{abcd}\int_k [D_S\fn{k}]_{cd} +4{\mathcal F}_{abcd}\int_k [D_P\fn{k}]_{cd}.
\label{selfPenergy}
}
Inserting Eq.\,\eqref{treeSpro}, Eq.\,\eqref{treePpro}, Eq.\,\eqref{selfSenergy} and Eq.\,\eqref{selfPenergy} into Eq.\,\eqref{fullSpro} and Eq.\,\eqref{fullPpro}, we have
\al{
[D_S^{-1}\fn{k}]_{ab}&= -k^2 \delta_{ab} +(M_S^2)_{ab},
\label{dressProDS}
\\
[D_P^{-1}\fn{k}]_{ab}&= -k^2 \delta_{ab} +(M_P^2)_{ab}.
\label{dressProDP}
}
The squared masses $(M_S^2)_{ab}$ and $(M_P^2)_{ab}$ are evaluated by the following self-consistent equations:
\al{
(M_S^2)_{ab}&=(m_S^2)_{ab}
+4{\mathcal F}_{abcd}\int_k[D_S\fn{k}]_{cd} 
+4{\mathcal H}_{abcd}\int_k[D_P\fn{k}]_{cd},
\label{SCMS}
\\
(M_P^2)_{ab}&=(m_P^2)_{ab}
+4{\mathcal H}_{abcd}\int_k[D_S\fn{k}]_{cd}
+4{\mathcal F}_{abcd}\int_k[D_P\fn{k}]_{cd}.
\label{SCMP}
}
Since $V_2$ does not depend on $\bar \sigma$, the gap equation is given by
\al{
\frac{\delta V_\text{eff}\fn{\bar \sigma,D_S,D_P;T}}{\delta {\bar \sigma}_a}=0\iff&
h_a=m^2{\bar \sigma}_a
-3{\mathcal G}_{abc}\left[ {\bar \sigma}_b{\bar \sigma}_c
+\int_k \left([D_S\fn{k}]_{ab} -[D_P\fn{k}]_{ab}  \right) \right]\nn
&\qquad + 4{\mathcal F}_{abcd}\left[ \frac{1}{3}{\bar \sigma}_b{\bar \sigma}_c
 + \int_k \left([D_S\fn{k}]_{ab}\right)\right]{\bar \sigma}_d\nn
&\qquad + 4{\mathcal H}_{bcad} {\bar \sigma}_d\int_k \left([D_S\fn{k}]_{cb}\right).
}
We write simply $V_\text{eff}\fn{\bar \sigma,D_S,D_P;T}=V_\text{eff}\fn{\bar \sigma,T}$.
By solving  the self-consistent equations Eq.\,\eqref{SCMS} and Eq.\,\eqref{SCMP}, we obtain the dressed masses $M_S$ and $M_P$ and the propagators Eq.\,\eqref{dressProDS} and Eq.\,\eqref{dressProDP}.
Using them, we can evaluate the effective potential Eq.\,\eqref{fTeffectivepotential}.

Note that the thermal integral appearing in Eq.\,\eqref{fTeffectivepotential} 
\al{
\int _k \log\fn{D^{-1}\fn{k}}=T\int\frac{\df^3k}{(2\pi)^3} \log\fn{k^2+m^2}
}
can be written as
\al{
T\int\frac{\df^3k}{(2\pi)^3} \log\fn{k^2+m^2} 
=\int\frac{\df^3k}{(2\pi)^3}\omega
+T\int\frac{\df^3k}{(2\pi)^3}\log\fn{1-e^{-\omega/T}},
\label{thermaleffects}
}
where $\omega=\sqrt{\vec k^2+m^2}$.
The first term corresponds to the one-loop effect at zero temperature.
This contribution is small~\cite{Lenaghan:1999si,Lenaghan:2000ey}, and then we ignore it.
The second term is the thermal loop correction and often is evaluated by the high temperature expansion or the fitting; see e.g.~\cite{Hamada:2016gux}.
In this work,  we numerically evaluate this integral.

\section{Formulation of Gravitational wave spectrum}\label{FGWsformula}
\subsection{The latent heat and the duration time}
As will be seen in the subsection~\ref{GWspectrum}, the GW background {spectra} are characterized by three quantities, namely, $T_t$, $\alpha$ and $\tilde \beta$.
We here briefly describe their definitions by following~\cite{Linde:1981zj,Linde:2005ht}.

Given the effective potential Eq.\,\eqref{fTeffectivepotential}, we first define the latent heat,
\al{
\Delta\epsilon\fn{T} =-V_\text{eff}\fn{\sigma_0\fn{T},T} + T \frac{\p V_\text{eff}\fn{\sigma_0\fn{T},T} }{\p T},
\label{latentheat}
}
where $\sigma_0\fn{T}=\bra \sigma\ket$ at $T$ is the expectation value in the broken phase.
The parameter $\alpha$ is defined as
\al{
\alpha = \frac{\Delta\epsilon\fn{T_t}}{\rho_\text{rad}\fn{T_t}},
\label{alphaparameter}
}
where $\rho_\text{rad}\fn{T}$ is the radiation energy density given by
\al{
\rho_\text{rad}=\frac{\pi^2}{30}g_*\fn{T}T^4,
}
and $T_t$ defined below is the cosmological phase transition temperature.
We see that the parameter $\alpha$ corresponds to the normalized latent heat at the phase transition. 

Next, in order to define the parameter $\tilde \beta$, let us start with defining the bubble nucleation rate which reads
\al{
\Gamma\fn{t} =\Gamma_0\fn{t}e^{-S_E\fn{t}}.
} 
Here $S_E\fn{t}=S_3\fn{T}/T$ is the three dimensional Euclidean action with
\al{
S_3\fn{T}&=\int \df^3x \left[ \frac{1}{2}(\p_i \sigma_B)^2 +V_\text{eff}\fn{\sigma_B,T} \right]\nn
&=4\pi \int \df r\,r^2 \left[ \frac{1}{2}\left( \frac{\df \sigma_B}{\df r}\right)^2 +V_\text{eff}\fn{\sigma_B,T} \right].
\label{eucideanaction}
}
Note that the time $t$ relates to temperature $T$ through $dT/dt=-HT$ where the Hubble parameter $H$ is
\al{
H^2\fn{T}=\frac{8\pi G_N}{3}\rho_\text{rad}\fn{T},
}
with the Newton constant $G_N$.
The field $\sigma_B\fn{r}$ is the bounce solution which satisfies the following equation of motion:
\al{
\frac{\df^2\sigma_B}{\df r^2} + \frac{2}{r}\frac{\df \sigma_B}{\df r} -\frac{\p V_\text{eff}}{\p \sigma_B}=0,
\label{EoMbounce}
}
with the boundary conditions
\al{
\frac{\df \sigma_B}{\df r}\bigg|_{r=0}&=0,&
\lim_{r\to \infty} \sigma_B&=\sigma_S.& 
\label{boundaryconditions}
}
Note that the effective potential is normalized as $V_\text{eff}\fn{\sigma_S,T}=0$ where $\sigma_S$ is the expectation value after the phase transition.
In the presence of the explicit breaking term, $\sigma_S$ could have a finite vacuum expectation value after the phase transition.
In the case without the explicit breaking term, we have $\sigma_S=0$ after the phase transition.
We here define the temperature $T_t$ (or the time $t_t$) at which the cosmological phase transition takes place.
The equation Eq.\,\eqref{EoMbounce} describes the phase transition from $\sigma_S$ to $\sigma_0$, that is, the dynamics for the bubble of the broken phase.
Since the universe is expanding, the bubble nucleation rate has to be compared with the Hubble time and volume.
Then at $T_t$ (or $t_t$) we have
\al{
\frac{\Gamma}{H^4}\bigg|_{T=T_t}\simeq 1.
}
It can be rewritten as
\al{
\frac{S_3\fn{T_t}}{T_t} 
=4\log\fn{T_t/H_t}
\simeq 140\text{--}150.
\label{phcriterion}
}
This is a criterion for the cosmological phase transition.

We now define the parameter $\beta$ by
\al{
S_E\fn{t}=S_E\fn{t_t} -\beta (t-t_t) +\cdots,
} 
with
\al{
\beta = -\frac{\df S_E}{\df t}\bigg|_{t=t_t}
=T_t\frac{\df}{\df T}\left( \frac{S_3\fn{T}}{T} \right)\bigg|_{T=T_t}
=\frac{1}{\Gamma} \frac{\df \Gamma}{\df t}\bigg|_{t=t_t}.
\label{betaparameter}
}
Hence, the parameter $\beta^{-1}$ is the duration time of the phase transition.
It is convenient to define the dimensionless duration time by
\al{
\tilde \beta= \frac{\beta}{H_t},
\label{tildebetaparameter}
}
where $H_t$ is the Hubble parameter at the phase transition, $H_t=H\fn{T_t}$.

\subsection{Formulas for {spectra} of gravitational waves}\label{GWspectrum}
Three processes of the gravitational waves due to the cosmological phase transition are known:
\al{
\Omega_\text{GW}\fn{\nu} {\hat h}^2\simeq \left[\Omega_\text{coll}\fn{\nu} + \Omega_\text{SW}\fn{\nu}  + \Omega_\text{MHD}\fn{\nu} \right] {\hat h}^2.
}
The first term is the spectrum generated by the collision of bubble walls~\cite{Kosowsky:1991ua,Kosowsky:1992rz,Kosowsky:1992vn,Kamionkowski:1993fg,Caprini:2007xq,Huber:2008hg,Jinno:2016vai} and is given by
\al{
\Omega_\text{coll}\fn{\nu}{\hat h}^2={\tilde \Omega}_\text{coll}{\hat h}^2 \left( \frac{\nu}{\tilde \nu_\text{coll}} \right)^{2.8}\frac{3.8}{1+2.8\left(\nu/\nu_\text{coll}\right)^{3.8}},
}
where 
\al{
{\tilde \Omega}_\text{coll}{\hat h}^2\simeq
1.67\times10^{-5}\left(\frac{0.11v_b^3}{0.42+v_b^2} \right){\tilde \beta}^{-2}\left(\frac{\kappa_\phi \alpha}{1+\alpha} \right)^2\left( \frac{100}{g_*^t}\right)^{1/3}
}
is the peak spectrum at the peak frequency
\al{
\tilde \nu_\text{coll}
\simeq 1.65\times 10^{-5}\, \text{Hz}\times \left(\frac{0.62}{1.8-0.1v_b+v_b^2}\right){\tilde \beta}\left( \frac{T_t}{100\,\text{GeV}}\right)\left(\frac{g_*^t}{100} \right)^{1/6}.
} 
Here $g_*^t=g_*\fn{T_t}$, $v_b$ is the wall velocity and $\kappa_\phi$ is the fraction of the vacuum energy transferred to the gradient energy of the scalar field.
The second term is the GW spectrum from the sound wave~\cite{Hindmarsh:2013xza,Hindmarsh:2015qta,Giblin:2013kea,Giblin:2014qia} which reads
\al{
\Omega_\text{SW}\fn{\nu}{\hat h}^2=\tilde \Omega_\text{SW}h^2\left( \frac{\nu}{\tilde \nu_\text{SW}}\right)^3
\left(\frac{7}{4+3(\nu/\tilde \nu_\text{SW})^2}\right)^{7/2},
}
with the peak spectrum 
\al{
\tilde \Omega_\text{SW}{\hat h}^2 \simeq 2.65\times 10^{-6}\left(\frac{\tilde \beta}{v_b} \right)^{-1}\left( \frac{\kappa_v \alpha}{1+\alpha} \right)^2\left(\frac{100}{g_*^t}\right)^{1/3},
}
at the peak frequency
\al{
\tilde \nu_\text{SW}\simeq 1.9\times 10^{-5}\, \text{Hz}\times \frac{\tilde \beta}{v_b} \left( \frac{T_t}{100\, \text{GeV}}\right) \left(\frac{g_*^t}{100} \right)^{1/6}.
}
Here $\kappa_v$ stands for the ratio of the latent heat transformed into the bulk motion of the fluid and is given as~\cite{Espinosa:2010hh}\footnote{
The behavior of $\kappa_v$ as a function of $v_b$ is shown in \cite{Hashino:2016rvx}.
}
\al{
\kappa_v\fn{\alpha,v_b}= \begin{cases}
\frac{c_s^{11/5}\kappa_A\kappa_B}{(c_s^{11/5}-v_b^{11/5}\kappa_B)+v_bc_s^{6/5}\kappa_A} & \text{for }v_b\lsim c_s\\
\kappa_B+(v_b-c_s)\delta\kappa + \frac{(v_b-c_s)^3}{(v_J-c_s)^3}[\kappa_C-\kappa_B-(v_J-c_s)\delta\kappa] & \text{for }c_s < v_b <v_J\\
\frac{(v_j-1)^3v_J^{5/2}v_b^{-5/2}\kappa_C\kappa_D}{[(v_J-1)^3-(v_b-1)^3]v_J^{5/2}\kappa_C +(v_b-1)^3\kappa_D}
& \text{for }v_J \lsim v_b,
\end{cases}
}
where $c_s=0.577$ is the velocity of sound, $v_J$ is the Jouguet detonations given by
\al{
v_J=\frac{\sqrt{2\alpha/3+\alpha^2}+\sqrt{1/3}}{1+\alpha},
}
and
\al{
\kappa_A&\simeq \frac{6.9\alpha}{1.36-0.037\sqrt{\alpha}+\alpha}v_b^{6/5},&
\kappa_B&\simeq \frac{\alpha}{0.017+(0.997+\alpha)^{2/5}},&\nn
\kappa_C&\simeq \frac{\sqrt{\alpha}}{0.135+\sqrt{0.98+\alpha}},&
\kappa_D&\simeq \frac{\alpha}{0.73+0.083\sqrt{\alpha}+\alpha}.&
}
$\delta \kappa$ is the derivative of $\kappa_v$ with respect to $v_b$ at $v_b=c_s$, which is approximately given by
\al{
\delta \kappa\simeq -0.9\log\fn{\frac{\sqrt{\alpha}}{1+\sqrt{\alpha}}}.
}
Note that when $v_b=c_s$, we have $\kappa_v\fn{c_s,\alpha}=\kappa_B$.

The last term originates from the magnetohydrodynamic (MHD) turbulence (turbulence of plasma) after the bubble wall collisions~\cite{Kosowsky:2001xp,Dolgov:2002ra,Caprini:2006jb,Gogoberidze:2007an,Kahniashvili:2008pe,Kahniashvili:2009mf,Caprini:2009yp,Kisslinger:2015hua} and is given as
\al{
\Omega_\text{MHD}\fn{\nu}{\hat h}^2
\simeq \frac{{\tilde \Omega}_\text{MHD}{\hat h}^2 }{\left( 1+8\pi \nu/h_t \right)}\left( \frac{\nu}{\tilde \nu_\text{MHD}}\right)^3 \frac{1}{\left(1+\nu/{\tilde \nu}_\text{MHD} \right)^{11/3}},
}
where $h_t$ is the redshifted inverse Hubble parameter whose value becomes
\al{ 
h_t=1.65\times 10^{-5}\,\text{Hz}\times \left( \frac{T_t}{100\, \text{GeV}} \right) \left( \frac{g_*^t}{100}\right)^{1/6}.
}
The peak spectrum and frequency are given by
\al{
\frac{{\tilde \Omega}_\text{MHD}{\hat h}^2}{\left( 1+8\pi {\tilde \nu}_\text{MHD}/h_t \right)} & \simeq \frac{3.35\times 10^{-4}}{\left( 1+8\pi {\tilde \nu}_\text{MHD}/h_t \right)}\left(\frac{{\tilde \beta}}{v_b}\right)^{-1} \left(\frac{\kappa_\text{MHD}\alpha}{1+\alpha}\right)^{3/2}\left( \frac{100}{g_*^t}\right)^{1/3},\\
{\tilde \nu}_\text{MHD}&\simeq 2.7\times 10^{-5}\, \text{Hz}\times \frac{{\tilde \beta}}{v_b}\left( \frac{T_t}{100\, \text{GeV}}\right)\left( \frac{g_*^t}{100}\right)^{1/6},
}
respectively.
The factor $\kappa_\text{MHD}$ is~\cite{Caprini:2015zlo}
\al{
\kappa_\text{MHD}=\varepsilon \kappa_v,
}
where $\varepsilon$ denotes the fraction of the turbulent bulk motion.
Following \cite{Caprini:2015zlo}, we use $\varepsilon=0.05$ in this work.
Note that the spectrum from the turbulence of plasma explicitly depends on the Hubble time since it produces the GWs for several Hubble times~\cite{Caprini:2009yp}.
The model-independent analysis has been performed in \cite{Grojean:2006bp}.

\section{The functions $B$ and $C$}\label{BC functions}
The amplitude for {pion-pion scattering} depends on the functions $B\fn{s,t,u}$ and $C\fn{s,t,u}$.
In {chiral perturbation theory}, they are expanded as
\al{
B\fn{s,t,u}&=B_\text{LO}\fn{s,t,u}+B_\text{NLO}\fn{s,t,u}+{\mathcal O}\fn{p^6},\\
C\fn{s,t,u}&=C_\text{LO}\fn{s,t,u}+C_\text{NLO}\fn{s,t,u}+{\mathcal O}\fn{p^6},
}
where the Mandelstam variables is defined in Eq.\,\eqref{Mandelstam variables stu}.
We here list these terms by following the literature~\cite{Bijnens:2011fm}.
To this end, we define 
\al{
x_2&=\frac{m_\pi^2}{f_\pi^2},&
L&=\frac{1}{16\pi^2}\log\fn{\frac{m_\pi^2}{\mu^2}},&
\pi_{16}&=\frac{1}{16\pi^2}.
}

For the lowest order, we have
\al{
B_\text{LO}\fn{s,t,u}&=x_2\left( -\frac{1}{2}t+1\right),&
C_\text{LO}\fn{s,t,u}&=0.
}
For the NLO, the functions are written as
\al{
B_\text{NLO}\fn{s,t,u}&=x_2^2\big[ B_P\fn{s,t,u}+B_S\fn{s,t-u}+B_S\fn{u,t-s}+B_T\fn{t}\big],\\
C_\text{NLO}\fn{s,t,u}&=x_2^2\big[ C_P\fn{s,t,u}+C_S\fn{s}+C_T\fn{t}+C_T\fn{u}\big],
}
where $B_P$ and $C_P$ are
\al{
B_P\fn{s,t,u}&=\alpha_1+\alpha_2 t +\alpha_3 t^2+\alpha_4(s-u)^2,\\
C_P\fn{s,t,u}&=\beta_1+\beta_2 s +\beta_3 s^2 +\beta_4 (t-u)^2.
}
For the breaking pattern $\SU\fn{N_f}_L\times \SU\fn{N_f}_R\to \SU\fn{N_f}_V$, we have
\al{
\alpha_1&=\frac{2}{N_f}+\frac{2}{N_f}\pi_{16}+16L_8^r+16L_0^r-\frac{2}{3}N_f L-\frac{5}{9}N_f\pi_{16},\\
\alpha_2&=-4L_5^r -16L_0^r +\frac{5}{12}N_fL + \frac{11}{36}N_f \pi_{16},\\
\alpha_3&=L_3^r +4L_0^r -\frac{1}{16}N_fL -\frac{1}{24}N_f\pi_{16},\\
\alpha_4&=L_3^r -\frac{1}{48}N_fL -\frac{1}{36}N_f\pi_{16},\\
\beta_1&=32(L_1^r-L_4^r +L_6^r)-\frac{2}{N_f^2}(L+\pi_{16}),\\
\beta_2&=16L_4^r -32L_1^r,\\
\beta_3&=-\frac{3}{8}L +2L_2^r +8L_1^r -\frac{3}{8}\pi_{16},\\
\beta_4&=2L_2^r -\frac{1}{8}L-\frac{1}{8}\pi_{16},
}
and
\al{
B_S\fn{s,t-u}&={\bar J}\fn{s}\left[ -\frac{1}{N_f} +\frac{N_f}{16}s^2 +\frac{N_f}{12}\left(1-\frac{s}{4}\right) (t-u) \right],\\
B_T\fn{t}&=0,\\
C_S\fn{s}&= {\bar J\fn{s}}\left( \frac{2}{N_f^2} +\frac{1}{4}s^2\right),\\
C_T\fn{t}&=\frac{1}{4}{\bar J}\fn{t} (t-2)^2,
}
where
\al{
{\bar J}\fn{s}=\pi_{16} (a ^2b +2),
}
with
\al{
a&=\sqrt{1-\frac{4}{s}},&
b&=\frac{1}{a}\log\fn{\frac{a-1}{a+1}}.&
}
In this work, we use ``$p^4$ fit" data given in table~1 of \cite{Bijnens:2014lea} for the coefficients $L_i^r$.
Note that we set $L_0^r$ to zero.
\end{appendix}


\bibliography{refs}

\end{document}